\pdfoutput=1
\newcommand*{\ATLASLATEXPATH}{}
\documentclass[PAPER, cernpreprint, atlasdraft=true, texlive=2016, UKenglish, LANGEDIT=true, LANGSHOW=false]{\ATLASLATEXPATH atlasdoc}
\LEcontact{Steve Lloyd s.l.lloyd@qmul.ac.uk}
 
\usepackage{\ATLASLATEXPATH atlaspackage}
\usepackage{\ATLASLATEXPATH atlasbiblatex}
 
\usepackage{\ATLASLATEXPATH atlasphysics}
 
\graphicspath{{logos/}{figures/}}
 
\usepackage{ANA-HDBS-2018-31-PAPER-defs}
 
\usepackage{fp}

\AtlasTitle{Search for diboson resonances in hadronic final states in 139 fb$^{-1}$ of $pp$ collisions at $\sqrt{s} = 13$ TeV with the ATLAS detector}
 
\AtlasAbstract{
Narrow resonances decaying into $WW$, $WZ$ or $ZZ$ boson pairs are searched for in 139 fb$^{-1}$  of proton--proton collision data at a centre-of-mass energy of $\sqrt{s}=13$ TeV recorded with the ATLAS detector at the Large Hadron Collider from 2015 to 2018. The diboson system is reconstructed using pairs of high transverse momentum, large-radius jets. These jets are built from a combination of calorimeter- and tracker-inputs compatible with the hadronic decay of a boosted $W$ or $Z$ boson, using jet mass and substructure properties. The search is performed for diboson resonances with masses greater than 1.3 TeV.
No significant deviations from the background expectations are observed.
Exclusion limits at the $95\%$ confidence level are set on
the production cross-section times branching ratio into dibosons for resonances in a range of theories beyond the Standard Model, with the highest excluded mass of a new gauge boson at 3.8 TeV in the context of mass-degenerate resonances that couple predominantly to gauge bosons.
}
 
\author{The ATLAS Collaboration}
 
\PreprintIdNumber{CERN-EP-2019-044}
 
\AtlasRefCode{HDBS-2018-31}

\AtlasCoverSupportingNote{Supporting Note}{http://cds.cern.ch/record/2647394}
 
\AtlasCoverCommentsDeadline{June 14th, 2019}
\AtlasJournal{JHEP}
\AtlasJournalRef{JHEP 09 (2019) 091}
\AtlasDOI{10.1007/JHEP09(2019)091}
\AtlasCoverAnalysisTeam{Sofia Adorni, Noemi Calace, Kalliopi Iordanidou, Roland Jansky, Enrique Kajomovitz Must, Bill Murray, Herjuno Rah Nindhito, Takuya Nobe, Yuta Okazaki, Steven Schramm}
 
\AtlasCoverEdBoardMember{Andrea Coccaro~(chair), Jeremy Love, Sven Menke}

\AtlasCoverEgroupEditors{atlas-HDBS-2018-31-editors@cern.ch}
 
\AtlasCoverEgroupEdBoard{atlas-HDBS-2018-31-editorial-board@cern.ch}

\hypersetup{pdftitle={ATLAS document},pdfauthor={The ATLAS Collaboration}}
 
\begin{document}
 
\maketitle

\section{Introduction}
\label{sec:introduction}
The discovery of new phenomena in high-energy proton--proton ($pp$) collisions is one of the main goals of the Large Hadron Collider (LHC). New heavy, \TeV{}-scale, resonances of vector bosons $VV$ (where $V$ represents a $W$ or a $Z$ boson) are a possible signature of such new physics and are predicted in several extensions to the Standard Model (SM).
These include extended gauge-symmetry models~\cite{Altarelli1989,EHLQ1984ref,Quigg2013},
Grand Unified theories~\cite{Pati:1974yy,Georgi:1974sy,Georgi:1974my,Fritzsch:1974nn}, theories with warped
extra dimensions~\cite{Randall:1999ee,KKmodesHLZ,ADPS2007,AAS2008,AS2008},
two-Higgs-doublet models~\cite{Branco:2011iw}, little-Higgs models~\cite{Perelstein2007247},
theories with new strong dynamics~\cite{1742-6596-259-1-012003},
including technicolour~\cite{Eichten:2007sx,Catterall:2011zf,Andersen:2011yj}, and more
generic composite Higgs models~\cite{Barducci:2013xaa}.
The data sample of 36.7~\ifb{} of $pp$ collisions collected in 2015 and 2016 at the LHC at $\sqrt{s}=13~\TeV{}$ offered improved sensitivity to heavy diboson resonances compared with earlier results.
The ATLAS and CMS collaborations performed searches in the fully hadronic final states using this data~\cite{EXOT-2016-19,CMS-B2G-17-001,Sirunyan:2019jbg} but no significant deviation from a smooth background consistent with the SM expectation was observed. Searches by ATLAS~\cite{EXOT-2016-28,EXOT-2016-29} and CMS~\cite{CMS-B2G-17-013,CMS-B2G-17-005} for semileptonic decay modes of the boson pair, as well as statistical combinations of various decay channels~\cite{EXOT-2017-31}, on the same data also did not reveal any hint of new physics.
 
This paper presents a search for narrow diboson resonances decaying into fully hadronic final states in $\lumi~\ifb$ of $pp$ collision data collected by the ATLAS experiment between 2015 and 2018. The $W$ and $Z$ bosons produced in the decay of \TeV{}-scale resonances are highly boosted, and are therefore reconstructed in ATLAS as a single large-radius-parameter jet. The signature of such heavy resonance decays is thus a resonant structure in the dijet invariant mass spectrum.
Although the hadronic decays of vector bosons have the largest branching ratio (67\% for $W$ and 70\% for $Z$ bosons), they suffer from background contamination from the production of multijet events.
This background is larger by several orders of magnitude, and to suppress it, the characteristic jet substructure of $W/Z$ boson decays is used.
Contributions to the background from SM processes containing bosons, $V+\mathrm{jets}$, SM $VV$, \ttbar and single top production, are significantly smaller.
 
To improve the sensitivity of this search, new techniques are used. Novel inputs are used for jet finding, which improve the jet substructure resolution of ATLAS in highly boosted topologies~\cite{ATL-PHYS-PUB-2017-015}. To further benefit from these developments, a new approach for identifying boosted boson candidates is introduced. The identification of the boosted boson candidates is validated using the known SM $V+\mathrm{jets}$ production.
 
To avoid limitations caused by poor modelling or limited numbers of Monte Carlo (MC) generated background events, the observed background is characterised by a parametric function fit to the smoothly falling dijet invariant mass distribution.
To assess the sensitivity of the search, to optimise the event selection and for comparison with the observed data, three specific benchmark models are used: a spin-0 radion~\cite{Oliveira:2014kla} decaying into $WW$ or $ZZ$; a spin-1 Heavy Vector Triplet (HVT) Model~\cite{Pappadopulo:2014}
that provides signals such as $W'\rightarrow WZ$ and $Z'\rightarrow WW$; and a spin-2 graviton $\grav\rightarrow WW$ or $ZZ$,
corresponding to Kaluza--Klein (KK) modes~\cite{Randall:1999ee,KKmodesHLZ} of the Randall--Sundrum (RS) graviton~\cite{ADPS2007,AAS2008,AS2008}. These models assume production mechanisms either via gluon--gluon fusion or quark--antiquark annihilation.
\section{ATLAS detector}
\label{sec:detector}
 
The ATLAS detector~\cite{PERF-2007-01} surrounds nearly the entire solid angle around the ATLAS collision point.
It has an approximately cylindrical geometry\footnote{ATLAS uses a right-handed coordinate system with its origin at the nominal interaction point (IP) in the centre of the detector and the $z$-axis along the beam pipe.
The $x$-axis points from the IP to the centre of the LHC ring, and the $y$-axis points upward. Cylindrical coordinates $(r,\phi)$
are used in the transverse plane, $\phi$ being the azimuthal angle around the beam pipe.
The pseudorapidity is defined in terms of the polar angle $\theta$ as $\eta=-\ln\tan(\theta/2).$ Angular distance is measured in units of  $\Delta R= \sqrt{(\Delta \eta)^2+(\Delta \phi)^2}$.} and consists of an inner tracking detector surrounded by
electromagnetic and hadronic calorimeters and a muon spectrometer.
The tracking detector is placed within a 2~T axial magnetic field provided by a superconducting solenoid and
measures charged-particle trajectories with silicon pixel and silicon microstrip detectors that cover the
pseudorapidity range $|\eta| < 2.5$,
and with a straw-tube transition radiation tracker covering $|\eta| < 2.0$. A new innermost pixel layer~\cite{Abbott:2018ikt,ibl} inserted at a radius of 3.3~cm has been used since 2015.
 
Electromagnetic and hadronic calorimeter systems provide energy measurements with high granularity.
The electromagnetic calorimeter is a liquid-argon (LAr) sampling calorimeter with lead absorbers,
spanning $|\eta| < 3.2$ with barrel and endcap sections.
The three-layer central hadronic calorimeter comprises scintillator tiles with steel absorbers and extends to $|\eta| = 1.7$.
The hadronic endcap calorimeters measure particles in the region $1.5 < |\eta| < 3.2$ using liquid argon with copper absorbers.
The forward calorimeters cover $3.1 < |\eta| < 4.9$, using LAr/copper modules for electromagnetic energy measurements
and LAr/tungsten modules to measure hadronic energy.
 
The muon spectrometer surrounds the calorimetry system and provides precision muon tracking and triggering.
It includes three large superconducting air-core toroids providing a magnetic field for accurate momentum measurements
in tracking drift chambers arranged in a barrel, covering $|\eta| < 1.0$, and endcaps, extending to $|\eta| = 2.7$. 
 
Events are recorded in ATLAS if they satisfy a two-level trigger requirement~\cite{TRIG-2016-01}.
The level-1 trigger detects jet and particle signatures in the calorimeter and muon systems with a fixed latency of 2.5~$\mu$s,
and is designed to reduce the event rate to about 100~kHz.
Jets are identified at level-1 with a sliding-window algorithm, searching for local maxima
in square regions with size $\Delta \eta \times \Delta \phi = 0.8 \times 0.8$.
The subsequent high-level trigger consists of software-based trigger filters that reduce the event rate to one~kHz.
\section{Data}
\label{sec:data}
 
The search is performed using data collected by the ATLAS experiment from 2015 to 2018 from $\sqrt{s}=13~\TeV{}$ LHC $pp$ collisions.
Events used in this search satisfied a single-jet trigger requiring at least one jet reconstructed at each trigger level.
The final filter in the high-level trigger required a jet to satisfy a high transverse momentum (\pt{})
threshold, $\pt\ge360~\GeV{}$ (2015), $\pt\ge420~\GeV{}$ (2016), $\pt\ge440~\GeV{}$ (2017 and 2018),
reconstructed with the \antikt{} algorithm~\cite{antikt} and a large radius parameter ($R=1.0$).
Calorimeter-cell energy clusters calibrated to the hadronic scale utilising the local cell signal weighting method~\cite{PERF-2014-07} were used as inputs. These are referred to as topo-clusters in this paper.
After requiring that the data were collected during stable beam conditions and the
detector components relevant to the analysis were functional,
the integrated luminosity was $\lumiFifteen~\ifb$ in 2015, $\lumiSixteen~\ifb$ in 2016, $\lumiSecenteen~\ifb$ in 2017 and $\lumiEightteen~\ifb$ in 2018.
\section{Simulation}
 
\subsection{Signal Models}
\label{sec:signalsamples}
MC simulation of signal events is used to optimise the sensitivity of the search
and to interpret its results. Signals are simulated in three benchmark scenarios.
 
In the first scenario, the gravitational fluctuations in the extra dimension of the Randall--Sundrum framework correspond to scalar fields, known as the radion, which are massless in the simplest scenario.
A fundamental problem in the original Randall--Sundrum framework is that it lacks a mechanism to stabilise the radius of the compactified extra dimension, $r_\mathrm{c}$.
One possible mechanism to achieve this is to introduce an additional bulk scalar \radion{}, produced via gluon--gluon fusion, which has its interactions localised on the two ends of the extra dimension~\cite{Goldberger:1999Modulus,goldberger2000phenomenology}. This causes the radion field to acquire a mass term, which is typically much smaller than the first KK excitation mass.
The coupling of the radion field to SM fields scales inversely proportional to the model parameter $\Lambda_\mathrm{R}=\sqrt{g}\times k \times \mathrm{e}^{-k\pi r_\mathrm{c}}\sqrt{M_5^3/k^3}$ where $M_5$ is the 5-dimensional Planck mass, which has been extensively studied in the literature~\cite{Oliveira:2014kla,barger2012randall,Csaki2007Radion,Csaki2007Radion}, $k$ the curvature factor, and $g$ is the 5-dimensional metric.
The size of the extra dimension, defined as $k\pi r_\mathrm{c}$, is another parameter of the model.
In this analysis, the curvature factor is set to $k\pi r_\mathrm{c}=35$, and $\Lambda_\mathrm{R}=3~\TeV{}$ is used.
 
The couplings of the \radion{} to fermions are proportional to the masses of the fermions, while the couplings are proportional to the square of the masses for bosons.
For \radion{} mass above $\sim1~\TeV{}$, the dominant decay mode is into pairs of bosons.
The decay width of the \radion{} is approximately 10\% of its pole mass, resulting in observable mass peaks with a width comparable to the experimental resolution (see Section~\ref{subsec:selefficiencies}). The calculated production cross-section times branching ratio (\sigbr{}) for a \radion{} decaying into $WW$, with the $W$ decaying hadronically, is 6.3~fb and 0.61~fb for \radion{} masses of $2~\TeV{}$ and $3~\TeV{}$, respectively. Corresponding values for a \radion{} decaying into $ZZ$ are 3.1~fb and 0.30~fb.
 
The second scenario is based on two benchmark models, A and B, of the HVT phenomenological Lagrangian~\cite{Pappadopulo:2014}. The Lagrangian introduces a new heavy vector triplet $V'$, where $V'$ refers to $W'$ and $Z'$, produced via quark--antiquark annihilation, whose members are degenerate in mass, and parameterises its couplings to SM fields in a generic manner, such that a large class of extensions to the SM can be described.
 
Model A with the strength of the vector-boson interaction $g_V = 1$~\cite{Pappadopulo:2014} describes scenarios where the new triplet field couples weakly to the SM fields and arises from an extension of the SM gauge group, with the heavy vector bosons having comparable branching ratios into fermions and gauge bosons.
For $W'$ and $Z'$ masses of interest, the width of the new heavy bosons is approximately 2.5\%, which results in observable mass peaks with a width dominated by the experimental resolution. The branching fraction of the new heavy boson $W'$ ($Z'$) to each of the $WZ$ and $WH$ ($WW$ and $ZH$) final states, where $H$ represents the Higgs boson, is approximately 2\%.
The calculated \sigbr{} values for $\wp \to WZ$, with $W$ and $Z$
bosons decaying hadronically, are 8.3~fb and 0.75~fb for $\wp$ masses of $2~\TeV{}$ and $3~\TeV{}$, respectively.
Corresponding values for $\zp \to WW$ are 3.8~fb and 0.34~fb.
 
Model B with $g_V = 3$ is representative of composite Higgs models, where the fermionic couplings to $V'$ are suppressed. The width of the resonances are identical to Model A. For the $W'$ and $Z'$ masses of interest,
the branching fraction of the new heavy boson $W'$ ($Z'$) to each of the $WZ$ and $WH$ ($WW$ and $ZH$) final states is close to 50\%.
Resonance widths and experimental signatures are similar to those obtained for model A and the predicted \sigbr{} values
for $\wp \to WZ$, with hadronic $W$ and $Z$ decays, are 13~fb and 1.3~fb for $\wp$ masses of $2~\TeV{}$ and $3~\TeV{}$, respectively.
Corresponding values for $\zp \to WW$ are 6.0~fb and 0.55~fb.
 
The third scenario considered is the bulk RS model~\cite{ADPS2007} that extends the original RS model~\cite{Randall:1999ee,Randall:1999vf} with a warped extra dimension, by
allowing the SM fields to propagate in the bulk of the extra dimension.
This model is characterised by a dimensionless coupling constant
$\kappa/\overline{M}_{{\mathrm{Pl}}}\sim1$, where $\kappa$ is the curvature of the warped extra dimension, and $\overline{M}_{{\mathrm{Pl}}}$ is the reduced Planck mass.
In this model, a Kaluza--Klein graviton, \grav{}, predominately produced via gluon--gluon fusion, decays into pairs of top quarks, pairs of Higgs bosons,  $WW$ and $ZZ$ with significant branching fractions.
The branching fraction of the \grav{} to $WW$ ($ZZ$) ranges from 24\% to 20\% (12\% to 10\%) as the mass increases. The decay width of the \grav{} is approximately 6\% of its pole mass, resulting in observable mass peaks with a width comparable to the experimental resolution, and \sigbr{} for $\grav \to WW$, with the $W$ decaying hadronically, is 1.29~fb and 0.06~fb for
$\grav$ masses of $2~\TeV{}$ and $3~\TeV{}$, respectively. Corresponding values for $\grav \to ZZ$ are 0.65~fb and 0.03~fb.
 
\subsection{Simulated event samples}
\label{sec:simulation}
 
For all MC samples, all hadronic decays were imposed at the generator level.
MC samples for the radion, HVT, and RS models, were generated using \MADGRAPHV{2.2.2}~\cite{madgraph}
interfaced to \PYTHIAV{8.186}~\cite{Sjostrand:2007gs} for hadronisation using the leading-order (LO) NNPDF 2.3 parton distribution function (PDF) set~\cite{Carrazza:2013axa} and the ATLAS A14 set of tuned parameters for the underlying event~\cite{ATL-PHYS-PUB-2014-021}. In all signal samples, the $W$ and $Z$ bosons are primarily longitudinally polarised.
The procedure to derive the optimal boson identification criteria (Section~\ref{sec:bosonidentification}) uses a dedicated sample of $W'$ decaying only into $W/Z$ bosons that in turn decay hadronically. \PYTHIAV{8.186} was used to generate this sample with the A14 set of tuned parameters for the underlying event and the NNPDF 2.3 LO PDF. The cross-section of the hard-scattering process was modified by applying an event-by-event weighting factor to broaden the width of the resonance and widen the \pt{} distribution of the electroweak bosons produced in its hadronic decays.
 
\PYTHIAV{8.186} with the NNPDF 2.3 LO PDF set
and the A14 set of tuned parameters was used to generate and shower multijet background events.
Samples of $W$+jets and $Z$+jets events were generated with
\SHERPAV{2.2.5}~\cite{Gleisberg:2008ta,Hoeche:2009rj,Gleisberg:2008fv,Schumann:2007mg} interfaced with the NNPDF 3.0 next-to-next-to-leading-order (NNLO) PDF set~\cite{ct10}. A \ttbar{} sample generated with \POWHEGBOX v2~\cite{Nason:2004rx,Frixione:2007vw,Alioli:2010xd} with the NNPDF 3.0 next-to-leading-order (NLO) PDF~\cite{Ball:2014uwa}, interfaced with \PYTHIAV{8.186} with the NNPDF 2.3 LO PDF and the A14 set of tuned parameters for parton showering is used for the $V$+jets study. EvtGen v1.2.0~\cite{LANGE2001152} was used for properties of bottom and charm hadron decays, except for samples generated by \SHERPA{}.
 
For all MC samples, the final-state particles produced by the generators were propagated through a detailed detector simulation based on {\textsc GEANT4}~\cite{geant,SOFT-2010-01}.
The mean number of $pp$ interactions per bunch crossing, `pile-up', was approximately
33 in the collision data being used for the analysis. The expected contribution from these minimum-bias $pp$ interactions was accounted for by overlaying additional minimum-bias events generated
with \PYTHIAV{8.186} using the ATLAS A3~\cite{ATL-PHYS-PUB-2016-017} set of tuned parameter. The MC simulation events were weighted to match the distribution of the average number of interactions per bunch crossing observed in collision data. Simulated events were then reconstructed with the same algorithms as run on the collision data.
\section{Reconstruction}
\label{sec:reconstruction}
 
The experimental signatures central to this analysis are hadronic jets.
Since the decay products of \TeV{}-scale resonances are highly boosted, their decay products become increasingly
collimated and they are therefore reconstructed as a single large-radius jet. It is important that they can still be differentiated from multijet events where a jet is initiated by a single quark or gluon. This relies on both the energy and angular resolution of the detector used to reconstruct the jet. Although the analysis primarily relies on jets, reconstruction of lepton candidates is necessary to reject events that could bias the SM $V$+jets studies presented in Section~\ref{sec:vplusjets}.
 
\subsection{\TCCsfull{}}
In previous analyses, ATLAS has mainly focused on the use of calorimeter-based jet substructure, which exploits the exceptional energy resolution of the ATLAS calorimetry~\cite{PERF-2014-07}. However, as the event becomes even more energetic, jets become so collimated that the calorimeter lacks the angular resolution to resolve the desired structure inside the jet. For boson jets with high transverse momentum, \pt{}, only a handful of calorimeter-cell clusters are created, each with limited angular resolution, but excellent energy resolution. On the other hand, the tracking detector has excellent angular resolution and good reconstruction efficiency at very high energy~\cite{PERF-2015-08}, while its momentum resolution deteriorates.
By combining information from the ATLAS calorimeter and tracking detectors, the precision of jet substructure techniques can be improved for a wide range of energies. This analysis uses a new unified object built from both the tracking and the calorimeter information, referred to as \TCCsfull{} (TCCs)~\cite{ATL-PHYS-PUB-2017-015}. This procedure is a type of particle flow, complementary to the energy subtraction algorithm described in the ATLAS particle flow paper~\cite{PERF-2015-09} which improves the energy resolution of low-energy jets. The two algorithms are designed to improve the jet reconstruction performance in very different energy regimes, reflected in their distinct four-momentum construction and energy sharing procedures. Energy sharing in the TCC approach is based solely on a weighting scheme where only the relative track momenta are used to spatially redistribute the energy measured in the calorimeter. In practice, this means that the TCC algorithm uses the spatial coordinates of the tracker and the energy scale of the calorimeter. A more detailed description of TCCs can be found in Ref.~\cite{ATL-PHYS-PUB-2017-015}.
 
\subsection{Jet reconstruction}
\label{sec:jetreconstruction}
This analysis uses \AKTFat{} jets reconstructed from both the combined and the neutral TCCs. Combined TCCs are four-momenta created by combining the angular information of tracks with the energy information of the calorimeters. Neutral TCCs are calorimeter topo-clusters that could not be matched to any track, most likely representing energy deposits from neutral particles. The use of combined and neutral TCCs captures most of the hard-scatter energy and provides the best representation of the total energy flow in the event, as there are both charged and neutral contributions. The combined TCC component is robust against effects from pile-up since only tracks consistent with coming from the primary vertex\footnote{If more than one vertex is reconstructed, the one with the highest sum of $\pt^2$ of the associated tracks is regarded as the primary vertex.} are used. However, by including the neutral TCCs, these jets  have a pile-up dependence similar to that of standard topo-cluster jets.  Jets are therefore trimmed~\cite{trimming} to remove contributions from pile-up by removing any $R=0.2$ subjet with less than 5\% of the \pt{} of the associated $R=1.0$ jet. The clustering and trimming algorithms use the FastJet package~\cite{Cacciari:2011ma}. The combination of pile-up suppression through track-to-primary-vertex matching and trimming makes these jets very robust against pile-up~\cite{ATL-PHYS-PUB-2017-015}.
A MC-based particle-level energy and mass calibration is applied to the jets, as described in Ref.~\cite{ATL-PHYS-PUB-2015-033}.
MC generator-level jets are built using the same algorithm and trimming procedure, with inputs of stable generator-level particles ($c \tau > 10\,\mathrm{mm}$) excluding muons and neutrinos, and excluding particles from pile-up. These serve as the reference in Figure~\ref{fig:perf}. The energy and mass of MC generator-level jets also serves as reference to which reconstructed detector-level jets are corrected to in the above mentioned calibration procedure. Consequently, the mass of jets from $V$ bosons is not expected to match the pole mass of the bosons.
 
Several jet properties can be used to discriminate hadronic decays of $W$ and $Z$ bosons from background jets. Two providing strong discrimination are the jet mass and \d2{},\footnote{The angular exponent $\beta$, defined in Ref.~\cite{Larkoski:2013eya}, is set to unity.} where the latter is defined as a ratio of two-point to three-point energy correlation functions that are based on the energies of the jets' constituents and their pairwise angular separations~\cite{Larkoski:2013eya}. Signal jets are expected to peak at \d2{} values below one, while jets from multijet background have significantly larger values. The radiation of a hard gluon can allow background jets to mimic a two-pronged structure and satisfy the tagging requirements described above. Discrimination between boson jets and multijet background from such gluon-initiated jets can be attained by selecting on the charged hadron multiplicity, in form of the track multiplicity (\ntrk{}) of the untrimmed $R=1.0$ jet, considering tracks with $\pt > 0.5~\GeV{}$ consistent with coming from the primary vertex.
 
Figure~\ref{fig:perf} shows the striking improvement in \d2{} resolutions\footnote{The resolution is defined as $R^{r} = [Q_{84}(\resp^{r}) - Q_{16}(\resp^{r})]/[2 \times Q_{50}(\resp^{r})]$ and $R^{d} = 1/2\left[Q_{75}(\resp^{d}) - Q_{25}(\resp^{d})\right]$ for the mass and \d2{}, respectively, where $Q_x$ is the $x\%$ quantile boundary, meaning that $Q_{50}$ is the median. The mass response is defined as $\resp^{r} = m_\mathrm{reco}/m_\mathrm{gen}$, while the residual of \d2{} is $\resp^{d} = D_{2,\mathrm{reco}} - D_{2,\mathrm{gen}}$, where `gen' and `reco' refer to the generated and reconstructed properties of the jets.} achieved with TCC jets. The mass resolution is superior to previously used jet mass variables ($m_{\textrm{comb}}$~\cite{JETM-2018-03}) starting around a jet \pt{} of $2~\TeV{}$. Below $1~\TeV{}$, the mass resolution in TCC jets is degraded. For identifying hadronically decaying $V$ bosons, the improvement in \d2{} resolution far outweighs the slight degradation in mass resolution.

\begin{figure}
\centering
\subfloat[]{\includegraphics[width=0.45\textwidth]{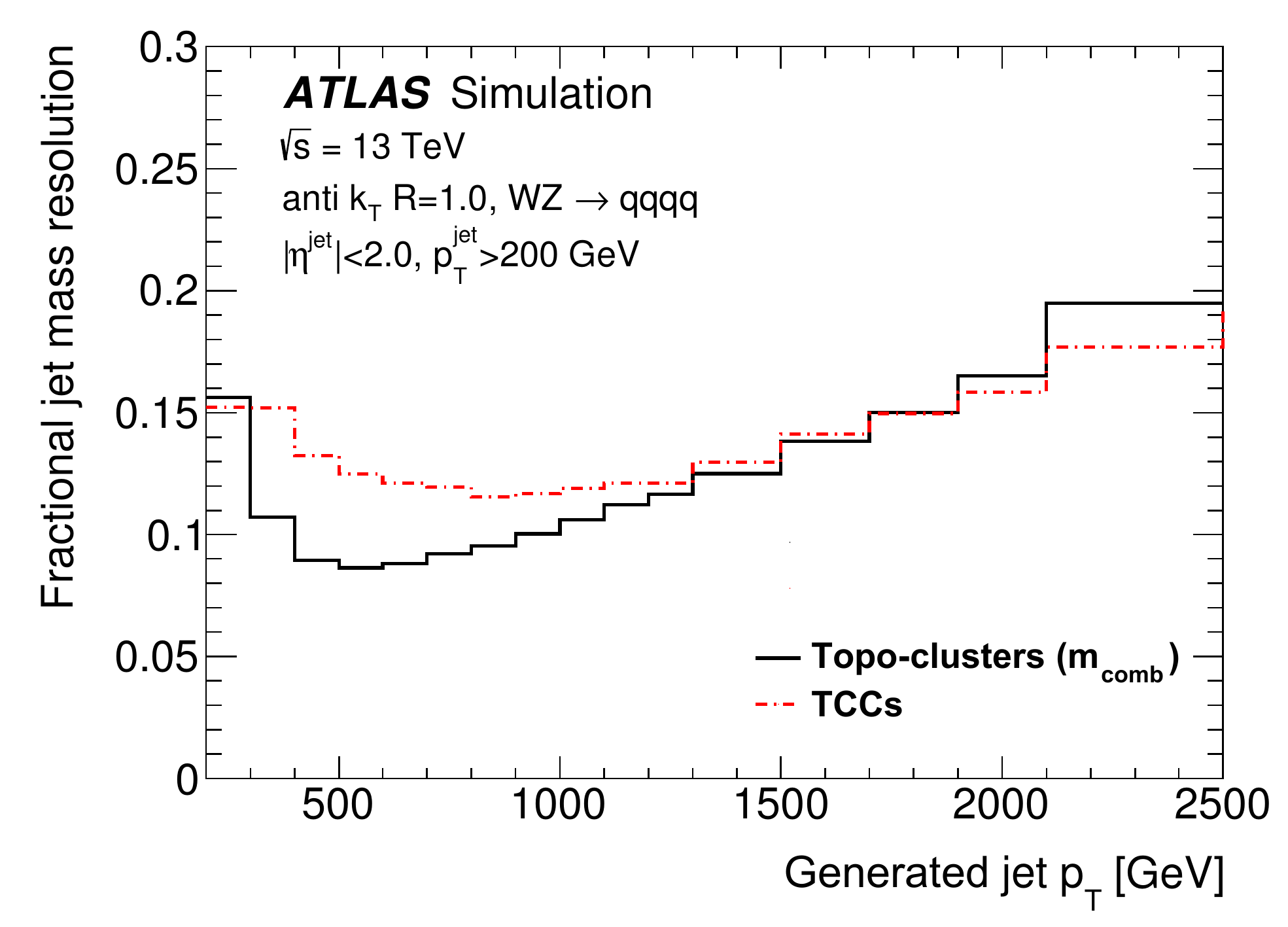}}
\subfloat[]{\includegraphics[width=0.45\textwidth]{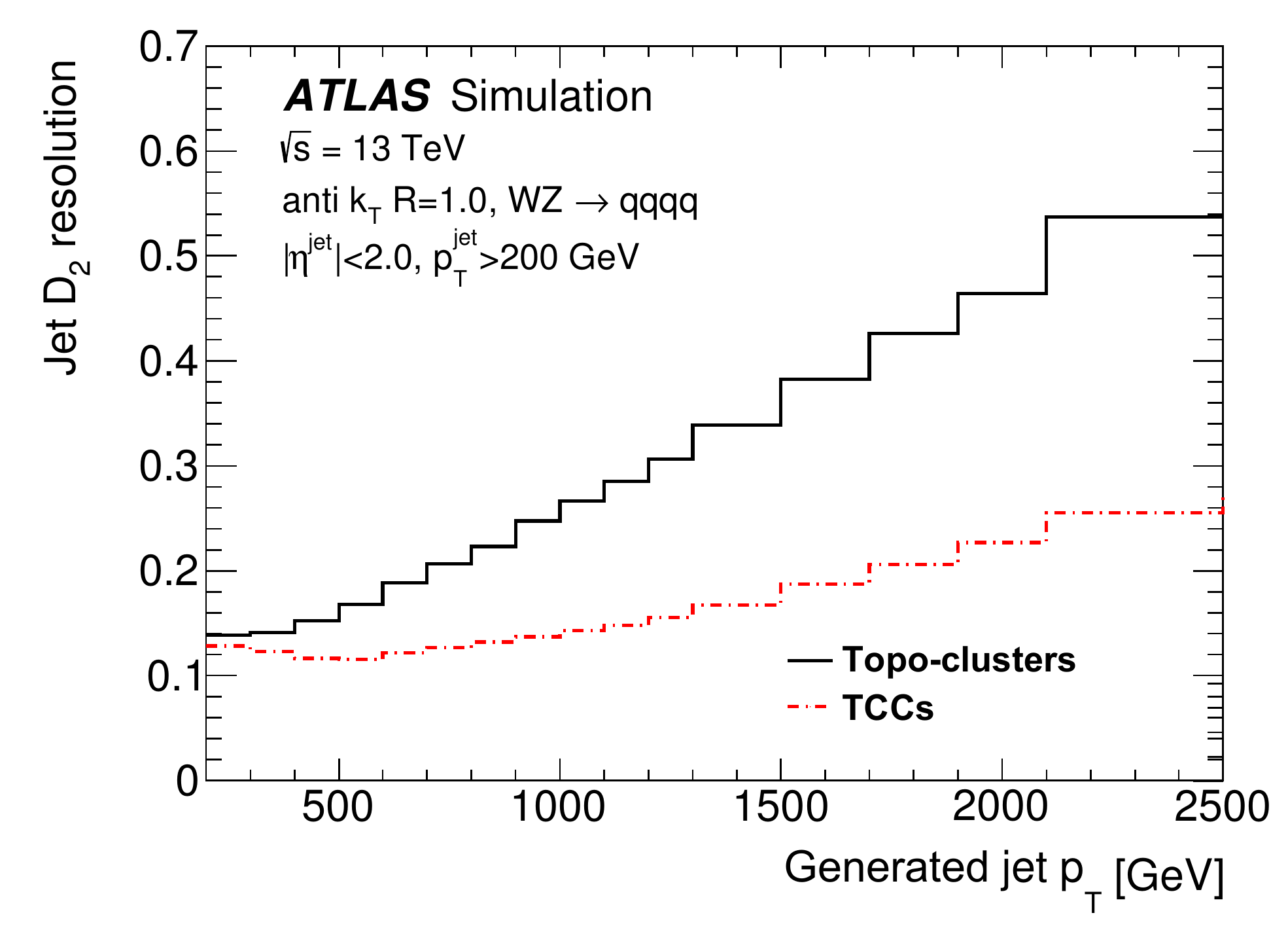}}
\caption{A comparison of (a) the fractional jet mass resolution for jets built from a linear combination of the calorimeter and track-only mass (Topo-clusters $m_{comb}$, solid line), and jets built using combined and neutral \TCCsfull{} objects (dashed lines) as a function of Monte Carlo generator-level jet \pt{}. The fractional jet resolution of the \d2{} variable (b) is compared between \TCCsfull{} and pure calorimeter jets. Only the two jets with the highest \pt{} per event matched to a generated jet from a $W$ or $Z$ boson are shown.}
\label{fig:perf}
\end{figure}
 
\subsection{Leptons}
\label{sec:leptons}
Electron identification is based on matching tracks to energy clusters
in the electromagnetic calorimeter and calculating a likelihood based on several properties of the electron candidate.
Electrons are required to have $\pt>25~\GeV{}$ and $|\eta| < 2.5$, and to satisfy the `medium' identification criterion~\cite{Aad:2019tso} and the `loose' track-based isolation~\cite{Aad:2019tso}.
 
Muon identification relies on matching tracks in the inner detector to muon spectrometer tracks or track segments.
Muons are required to have $\pt > 25~\GeV{}$ and $|\eta| < 2.5$, and to satisfy the `loose' selection criterion~\cite{PERF-2015-10} and the `loose' track isolation~\cite{PERF-2015-10}.
\section{Event selection}
\label{sec:selection}
 
To avoid contamination from non-collision backgrounds such as from calorimeter noise, beam halo, and
cosmic rays, events containing an \antikt{} jet built from calorimeter-cell clusters with $R=0.4$ and $\pt>20~\GeV{}$
failing to meet the loose quality criteria for consistency with production in $pp$
collisions are rejected~\cite{ATLAS-CONF-2015-029}. In addition, events with at least one lepton meeting the requirements defined in Section~\ref{sec:leptons} are rejected. There are no further requirements on leptons that are aligned with jets.
 
Events are required to have at least two \AKTFat{}, jets originating from the primary vertex, one with $\pt>500~\GeV{}$ and the second with $\pt>200~\GeV{}$. The leading (highest \pt{}) and subleading of these jets must satisfy $|\eta| < 2.0$ (to guarantee a good overlap with the tracking acceptance), have masses $m_\mathrm{J}>50~\GeV{}$, and their invariant mass, \mjj{}, must be larger than $\minmjj~\TeV{}$. The last requirement ensures that the triggers in use are fully efficient for the backgrounds and the benchmark signals. These selections are referred to as pre-selections.
 
The pair of jets is then required to have a small separation in rapidity,
$|\Delta y_{12}|<1.2$. This requirement reduces the multijet background,
which is mainly produced in $t$-channel processes with large
rapidity differences, in contrast to signal events that are expected to be produced in $s$-channel processes with small rapidity differences.
Additionally, to reject events with potentially badly reconstructed jets, a criterion is applied to the \pt{} asymmetry, $A = (\ptone - \pttwo) \, / \, (\ptone + \pttwo) < 0.15$, where $\ptone$
and  $\pttwo$ are the transverse momenta of the leading and
subleading jets, respectively.
 
\subsection{Vector-boson identification}
\label{sec:bosonidentification}
 
Jet substructure can be exploited to enhance the separation between signal boson jets and jets from multijet background.
Several promising variables have been studied~\cite{ATL-PHYS-PUB-2015-033}, with the largest sensitivity gain coming from the use of the three variables introduced in Section~\ref{sec:jetreconstruction}: jet mass, \d2{}, and \ntrk{}.
 
A three-dimensional (jet mass, \d2{}, \ntrk{}) tagger using TCC jets is optimised to provide maximum significance for boosted vector-boson jets relative to background jets.
A measure of significance, independent of the cross-sections of the new processes being searched for, is selected: $\epsilon/(a/2+\sqrt{B})$, where $\epsilon$ is the per-signal-jet selection efficiency for masses in the range of $0.5~\TeV{}$ to $10~\TeV{}$ in the $W'$ model described in Section~\ref{sec:signalsamples}, $a$ is the number of standard deviations corresponding to a one-sided Gaussian distribution, and $B$ is the number of background jets after the selection~\cite{Punzi:2003bu} taken from MC simulation. This number does not rely on a specific signal, but is valid for all signals with similar experimental features. Compared with the often used $S/\sqrt{B}$, that breaks down for small values of $B$, as is the case here, this measure is more appropriate. A value of $a=3$ is used, where the result of the optimisation is not very sensitive to the exact value.
For each jet \pt{} bin, the optimal selection on the three variables is defined by the combination of selections that leads to the highest significance. This simultaneous treatment properly accounts for the correlations between the variables. The result of this optimisation does not depend on the pre-selections described at the beginning of this section. Next, the applied selection criteria on jet mass, \d2{}, and \ntrk{} are parameterised with jet-\pt{}-dependent functions. For the jet mass, the function follows its approximate experimental resolution. For the latter two a simple higher-order polynomial is used.
The resulting smooth selections for the $W$ and $Z$ boson taggers as a function of jet \pt{} are shown in Figure~\ref{fig:massd2cuts}. Jets selected in the analysis are required to have jet masses inside the jet mass window and \d2{} and \ntrk{} values below the shown values. It should be noted that the $W$ and $Z$ boson mass windows overlap. This search is not sensitive to signatures containing massive particles with masses different from $W/Z$ boson masses (for example top quarks or Higgs bosons).
 
\begin{figure}
\centering
\subfloat[]{\includegraphics[width=0.45\textwidth]{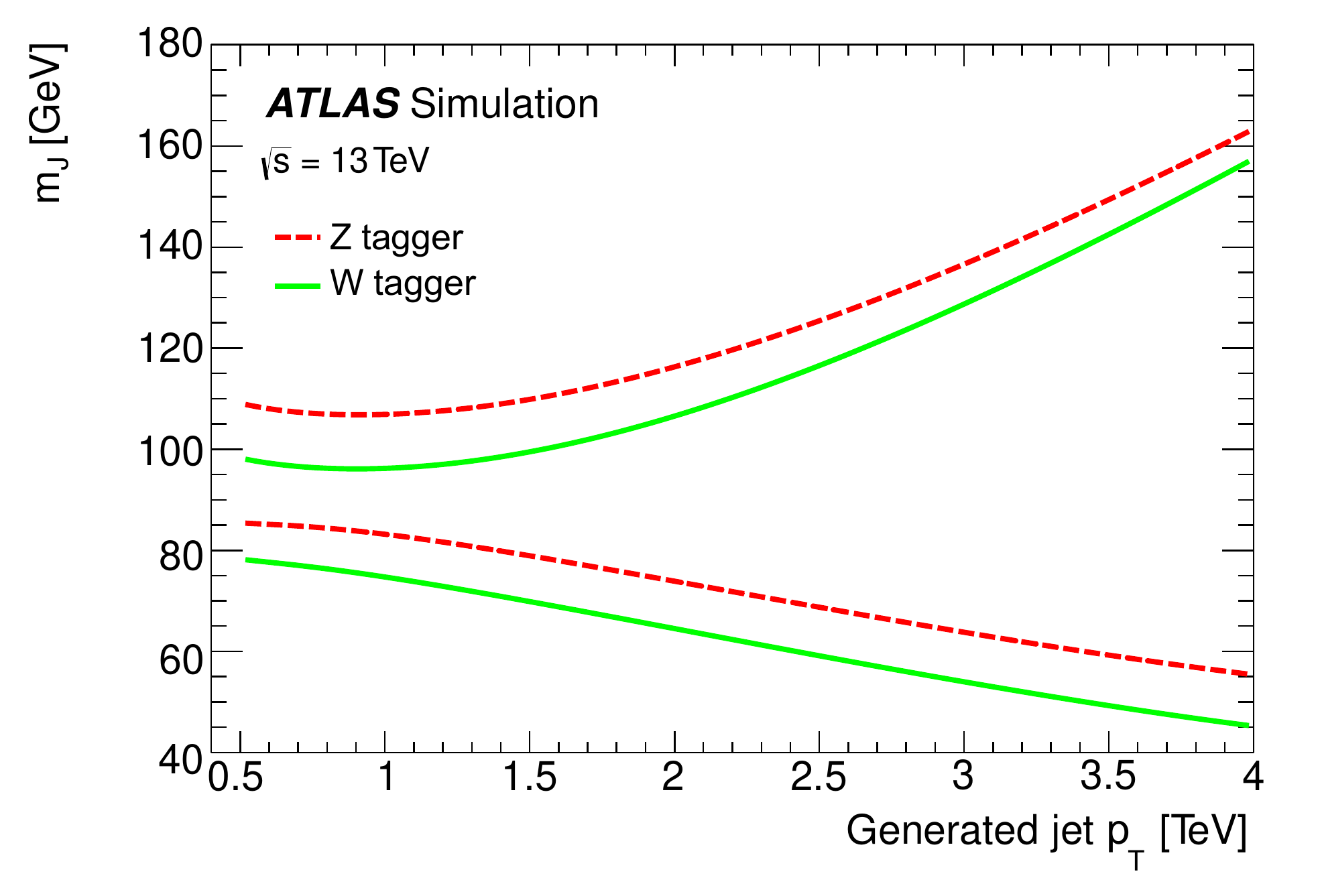}}
\subfloat[]{\includegraphics[width=0.45\textwidth]{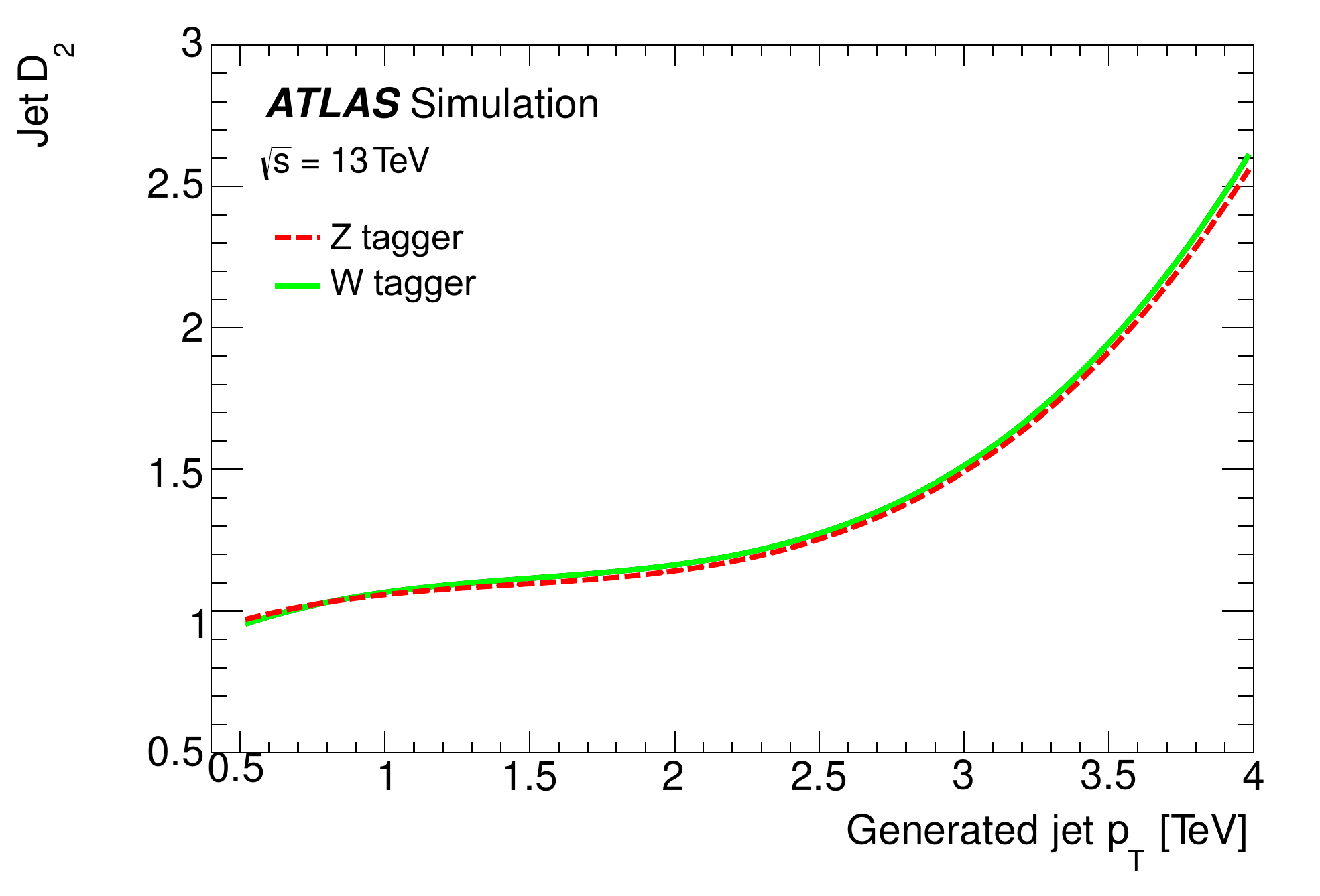}} \\
\subfloat[]{\includegraphics[width=0.45\textwidth]{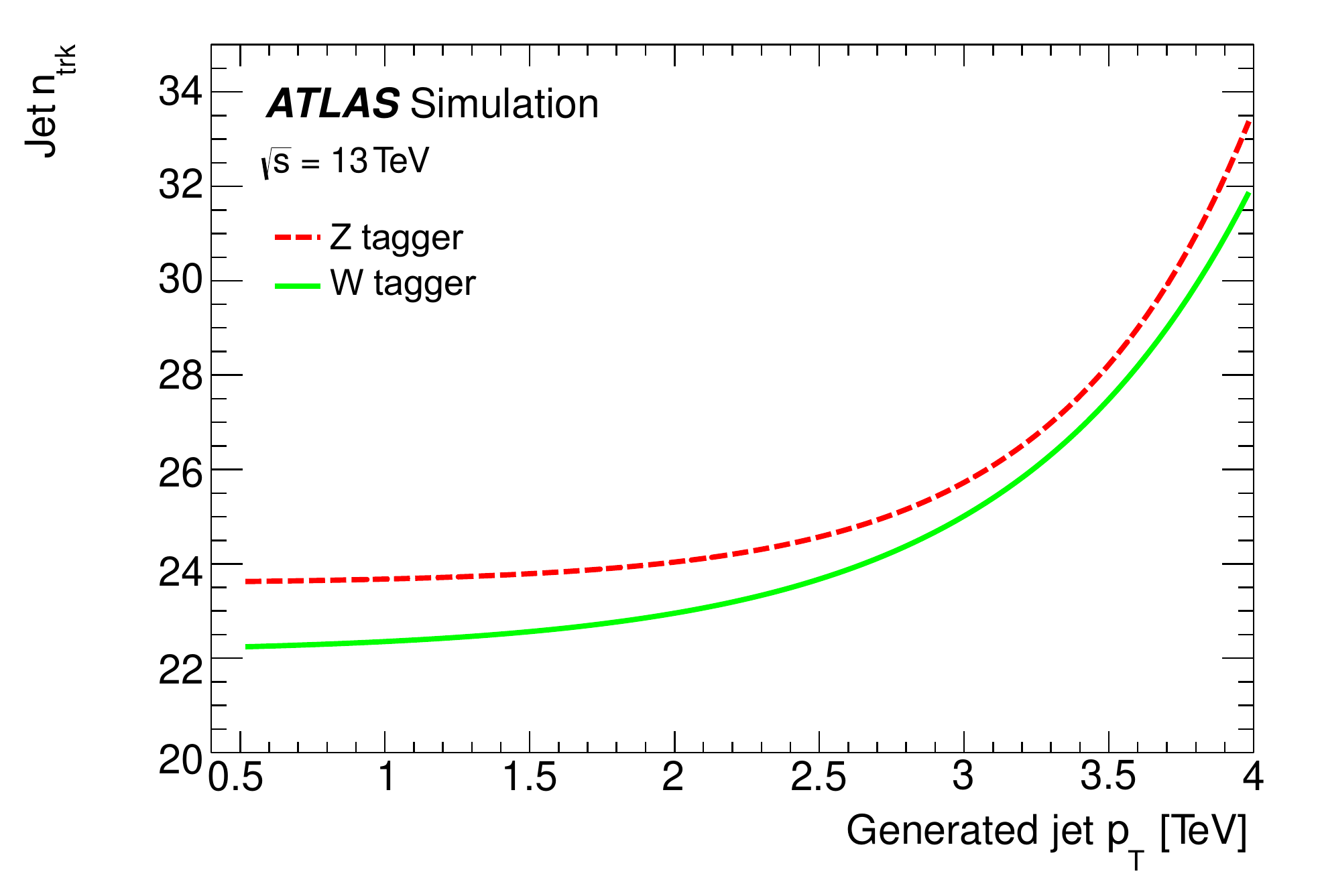}}
\caption{(a) Jet mass window, (b) \d2{} selection and (c) \ntrk{} selection  of the $W$ and $Z$ taggers as a function of jet \pt{}. Jets selected in the analysis are required to have jet masses inside the jet mass window and \d2{} and \ntrk{} values below the shown values. The tagger is only valid for jets with a \pt{} between $0.5~\TeV{}$ and $4.0~\TeV{}$ and with $|\eta^{\textrm{jet}}|<2.0$.}
\label{fig:massd2cuts}
\end{figure}
 
Unlike previous boson taggers~\cite{EXOT-2016-19}, the optimisation described above does not enforce a fixed signal efficiency nor a fixed background rejection, but rather creates a smooth behaviour that maximises the analysis sensitivity. Figure~\ref{fig:signaleffbkgrej} shows the resulting $W$ and $Z$ boson efficiencies and multijet background rejections (defined as $1/\textrm{efficiency}$) as a function of jet \pt{}. The selection criteria retain about 20\% efficiency for $W$ and $Z$ boson jets with $\pt=0.5~\TeV{}$. At higher jet \pt{}, the signal efficiency increases, reaching an efficiency close to 60\% for jets with a \pt{} of 4~\TeV{}. This is due to the behaviour of the multijet background, which  decreases rapidly for high dijet masses, and thus higher jet \pt{}. In the regime of $\mjj>3.0~\TeV{}$, where the number of background events is small, the tagger maintains a reasonable acceptance for signals with small cross-sections. This is also reflected in the background rejection as a function of jet \pt{}.
 
\begin{figure}
\begin{center}
\subfloat[]{\includegraphics[width=0.45\textwidth]{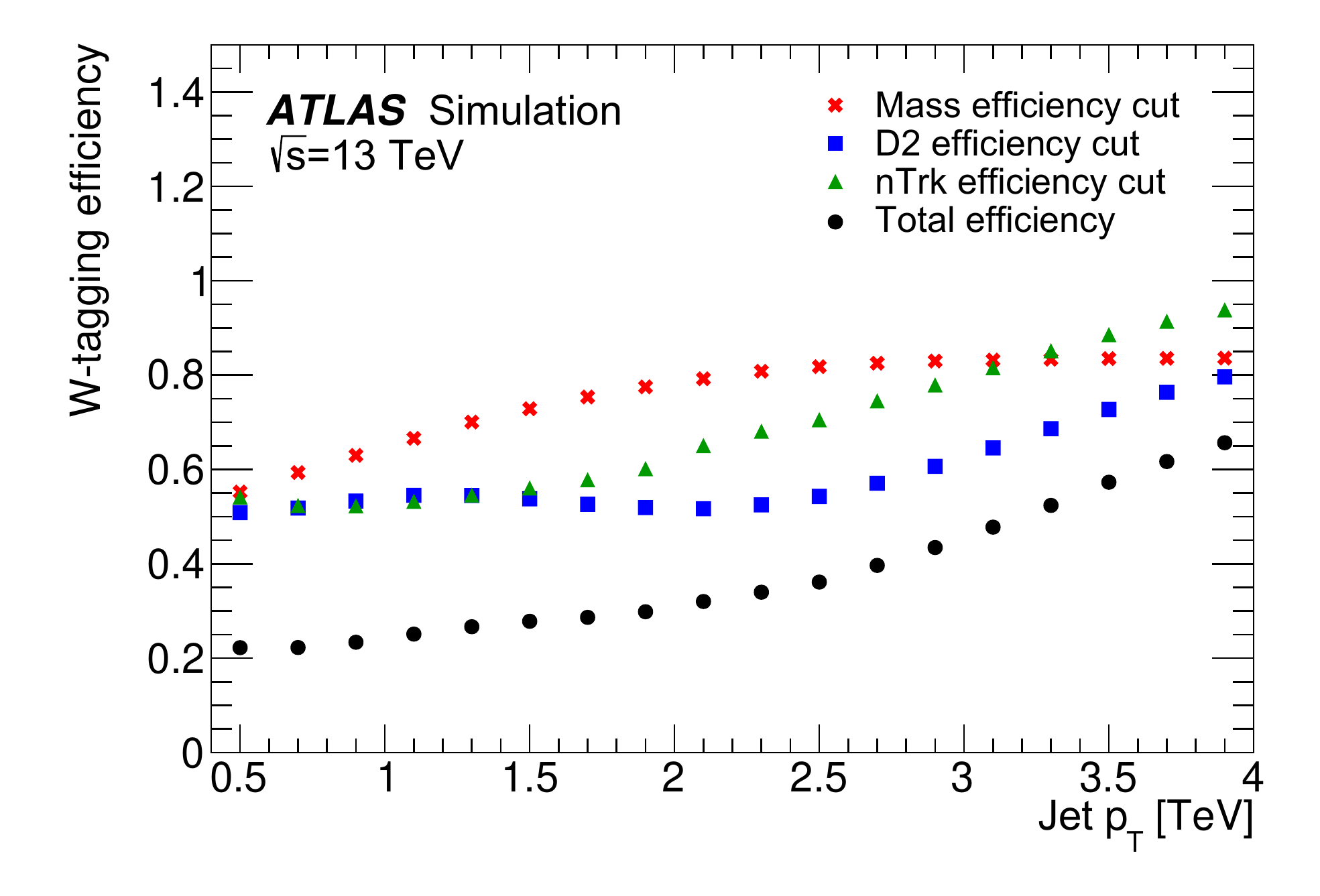}}
\subfloat[]{\includegraphics[width=0.45\textwidth]{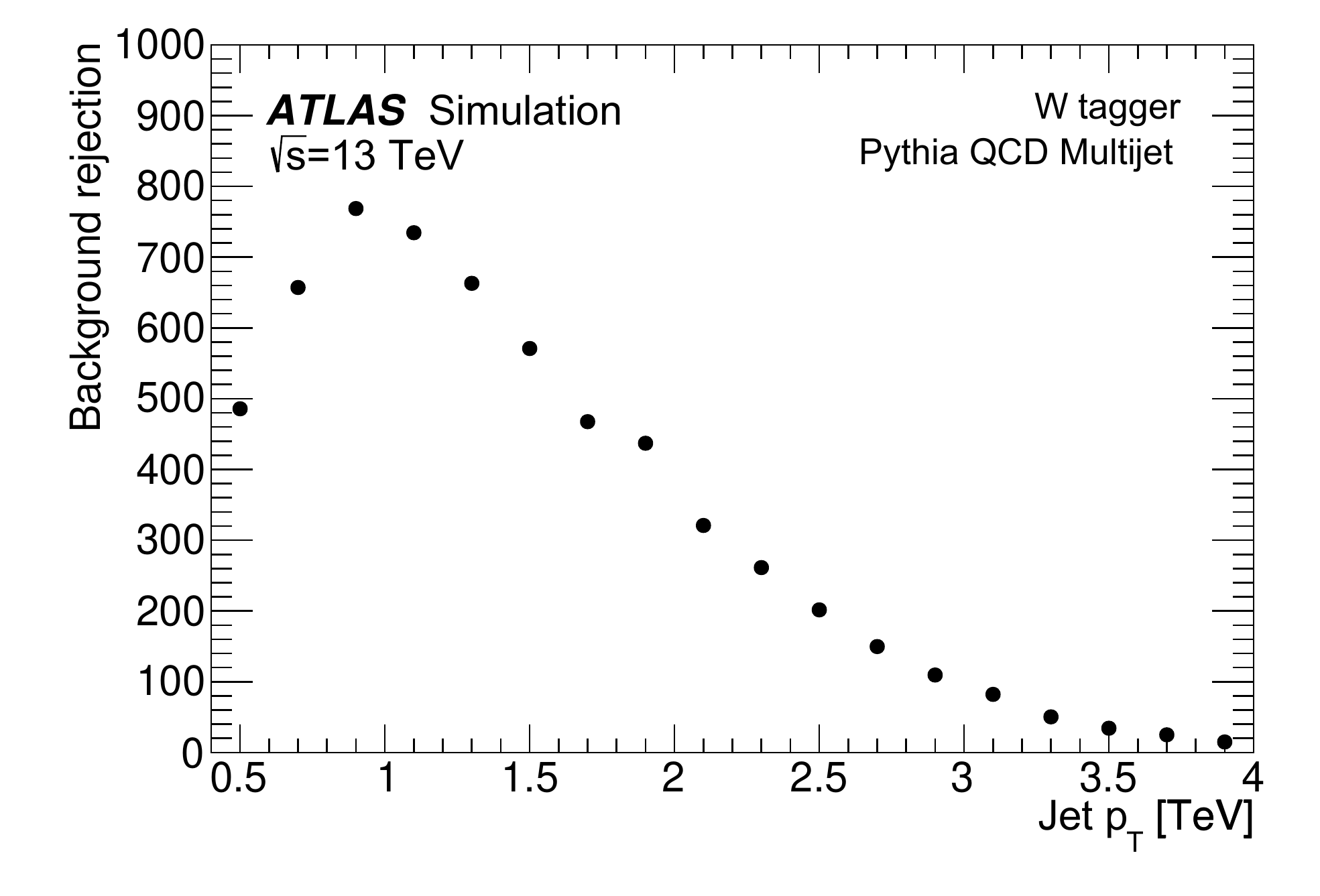}}
 
\subfloat[]{\includegraphics[width=0.45\textwidth]{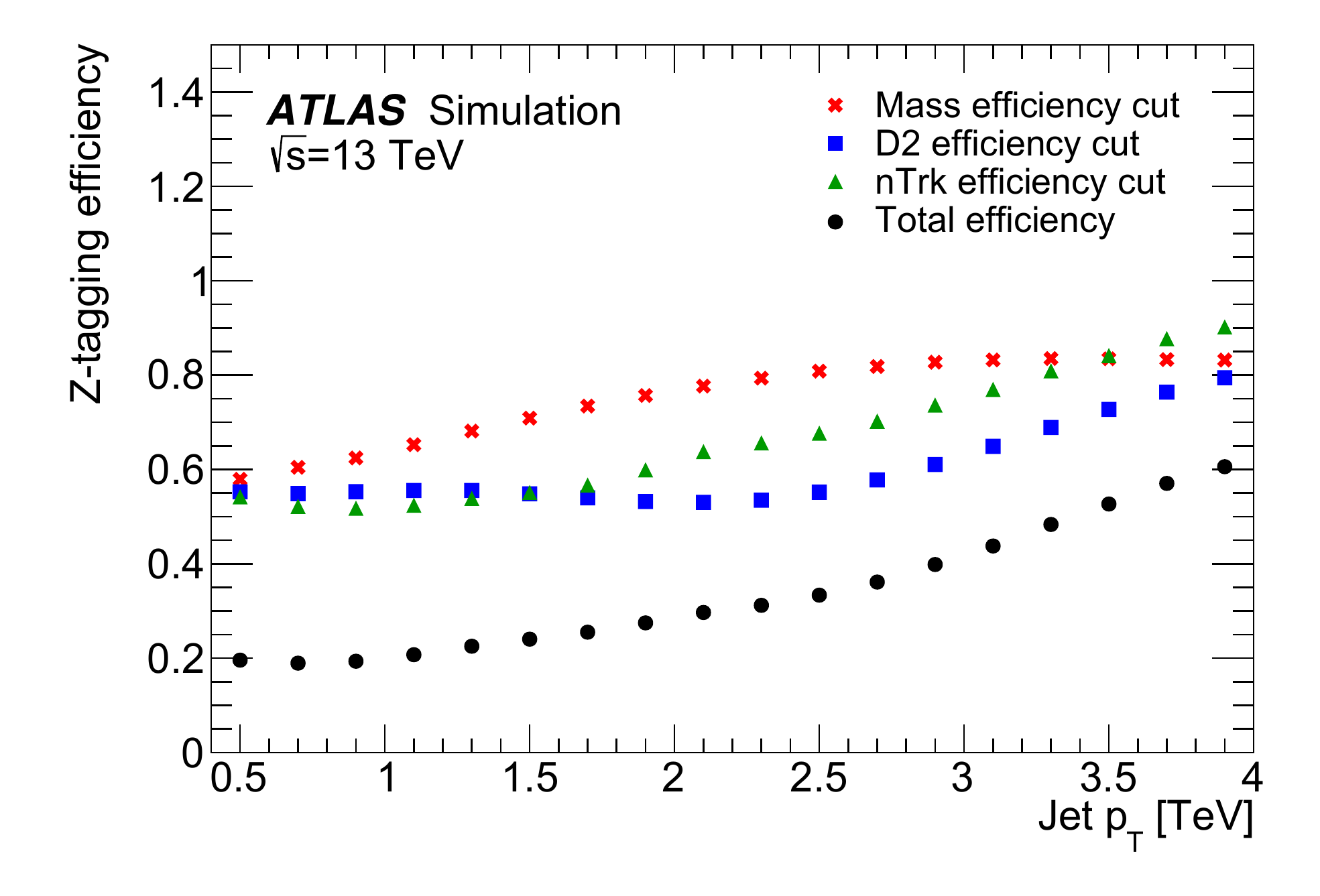}}
\subfloat[]{\includegraphics[width=0.45\textwidth]{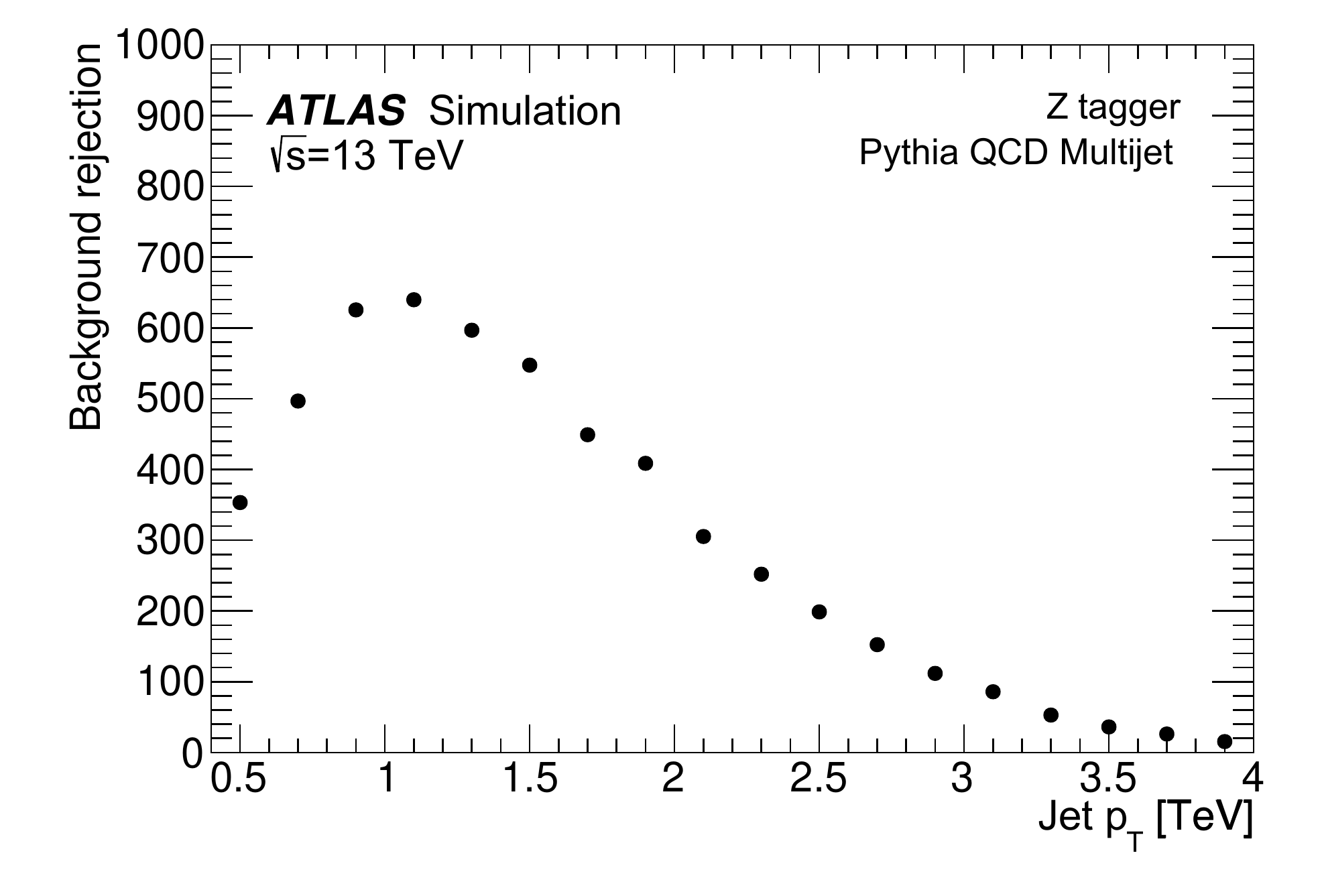}}
\caption{The (a) per-boson signal efficiency for the jet mass, \d2{}, and \ntrk{} selections, as well as the combined efficiency and (b) background rejection ($1/\textrm{efficiency}$) of the $W$ tagger for HVT $W' \rightarrow WZ \rightarrow qqqq$ and MC simulated multijets as a function of the jet \pt{}. Corresponding values for the $Z$ tagger are shown in (c) and (d).}
\label{fig:signaleffbkgrej}
\end{center}
\end{figure}
 
\subsection{Measurement of boson-tagging efficiency}
\label{sec:vplusjets}
The modelling of the boson-tagging efficiency is evaluated in a data
sample enriched in final states with a vector boson plus jets. This
sample is obtained by requiring two large-radius jets with $|\eta|<2.0$ and then
requiring that the leading jet has $\pt>600~\GeV{}$. A higher minimum \pt{} requirement is imposed on the leading jet than in the nominal event selection to obtain a sample with higher average leading jet \pt{} that better corresponds to the jet \pt{} values probed in the search. Events with identified leptons are vetoed.
Both jets are independently analysed  for the presence of a vector
boson by requiring them to satisfy the \d2{} and \ntrk{} selection for either a $W$ or a $Z$ boson.
The opposite jet is required to not satisfy the same \d2{} selection to
guarantee independence of this control region from the main analysis signal region.
 
The mass distribution of the selected jets between $50~\GeV{}$ and
$200~\GeV{}$ is fit by a signal-plus-background function, allowing the
inclusive rate of $V+\mathrm{jets}$ events to be measured. 
The contribution originating from $V+\mathrm{jets}$ processes is modelled using
a double-Gaussian distribution with the shape parameters determined from simulation,
while the background contribution is fit to data using a fourth-order exponentiated polynomial.
The relative rates of $W+\mathrm{jets}$ and $Z+\mathrm{jets}$ events in the fit are fixed from MC simulation.
The ability of the fit to extract the correct $V+\mathrm{jets}$ yield (also called MC closure) is tested in simulation by injecting signals of various strengths. Good linearity is found and the method is deemed reliable. By comparing the measured event yield in data and MC simulation,
potential differences in the selection efficiency ($s_\mathrm{Tag}$) can be probed. Expected contributions of about 5\% from $t\overline{t}$ events are subtracted based on MC simulation.
The cross-section of $V+\mathrm{jets}$ at a $V$ \pt{} of about $600~\GeV{}$ is modelled with about 10\% accuracy by the simulation~\cite{Lindert:2017olm}.
Additional systematic uncertainties in the fitted $V+\mathrm{jets}$ event yield from closure, from the uncertainty in the $t\overline{t}$ contribution, as well as from the
fit parameterisation are considered.
The relative efficiency of the \d2{} and \ntrk{} selections is extracted for $V$ bosons with \pt{} starting from
$600~\GeV{}$, while the analysis extends to $\pt=3.5~\TeV{}$. To estimate the dependence of the modelling on the jet \pt{}, the distribution of the \d2{} and \ntrk{} variables is compared in an inclusive sample in data and MC simulation as a function of jet \pt{}. The observed residual mismodelling as a function of jet \pt{} is taken into account as an additional 5\% uncertainty in the relative efficiency.
 
The fit to data is shown in Figure~\ref{Vj:data}.
This fit only extracts the overall yield, while the width and mean of the $W/Z$ peaks are fixed from similar fits performed on MC simulation. The fitted relative efficiency of the \d2{} and \ntrk{} selections in data compared with MC simulation is $s_\mathrm{Tag} =
0.92\pm0.04\ (\mbox{stat})\pm0.02\ (\mbox{closure})\pm 0.03\ (t\overline{t})\pm
0.02\ (\mbox{fit})\pm 0.05\ (\mbox{\pt range})\pm 0.10\ (\mbox{theory})$, or
$s_\mathrm{Tag} = 0.92\pm0.13$. This is applied as a scale-factor to the signal MC events, where the uncertainty in it reflects the uncertainty in the $W/Z$-tagging efficiency in the simulation. Additional fits allowing both the width and the mean of the $W/Z$ peaks to float are used to compare the efficiency of the jet mass window of the boson taggers in data and simulation. Excellent agreement is found, and no additional uncertainty is assigned. The polarisation of the vector bosons in this control region will be different from those in different signal models, but as no other physics process has sufficient sensitivity to probe the data-to-simulation agreement it is assumed that the simulation models the polarisation effects sufficiently well that the scale-factors can be applied globally.
 
\begin{figure}[htp]
\begin{center}
\includegraphics[scale=0.5]{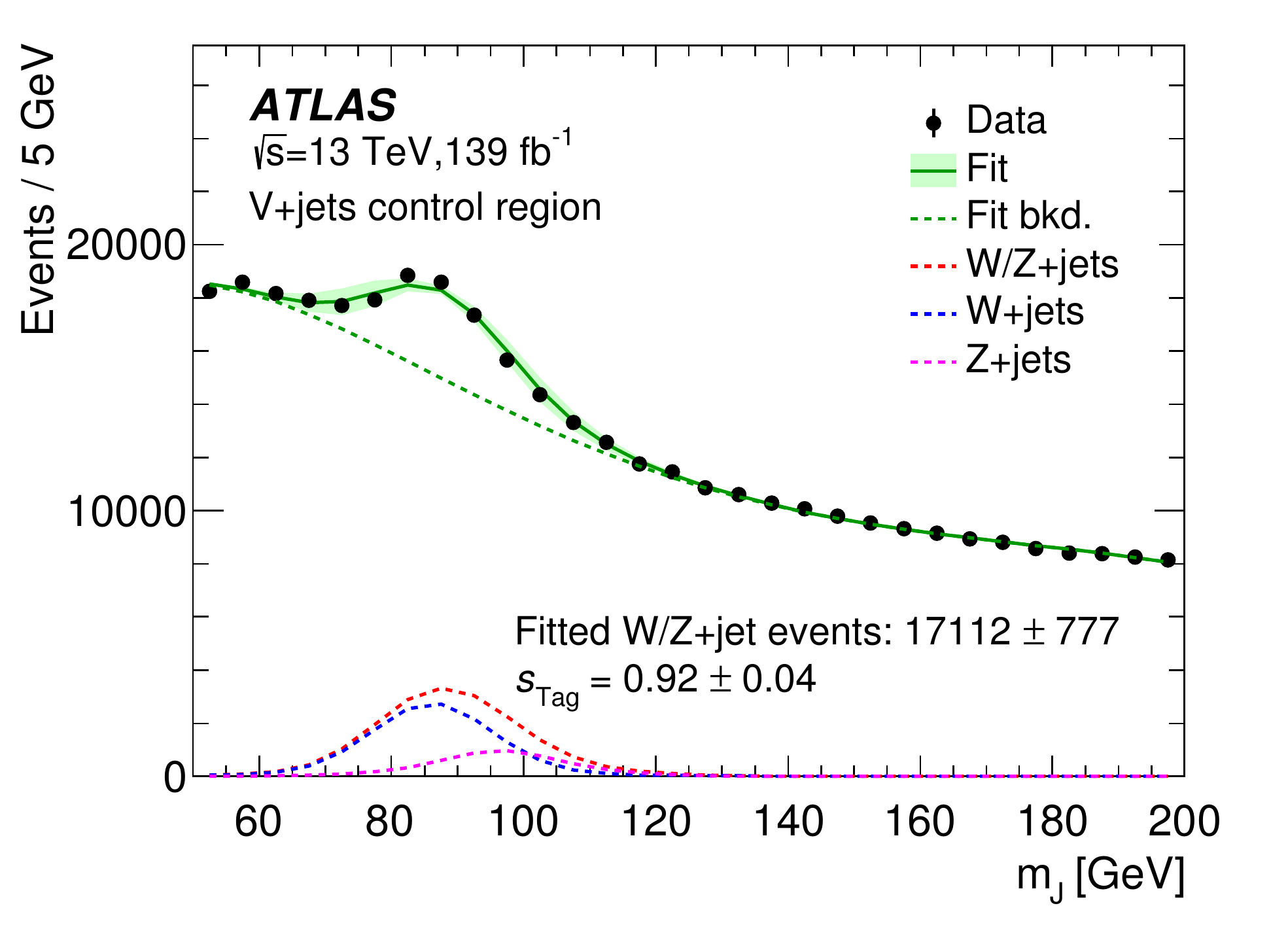}
\end{center}
\caption{
Jet mass distribution for data in the region enhanced in $V+\mathrm{jets}$ events  after boson tagging based only on the \d2{} and \ntrk{} variables.
The result of fitting to the sum of functions for the $V+\mathrm{jets}$ and background events is also shown. The shown fit uncertainty reflects the uncertainty in shape and positions of the $W$ and $Z$ peaks.  At the bottom, the fitted contribution to the observed jet mass spectra from the $V+\mathrm{jets}$ signal is shown. The fitted relative efficiency of the \d2{} and \ntrk{} selections is $s_\mathrm{Tag}=0.92\pm0.04$, where the uncertainty is purely statistical.
}
\label{Vj:data}
\end{figure}
 
\subsection{Signal and background selection efficiency}
\label{subsec:selefficiencies}
After boson tagging, the data is categorised into five non-exclusive signal regions (SRs): events with two jets identified as $WW$, $ZZ$, or $WZ$ form three SRs, and events with two jets identified as either $WZ$ or $WW$, and either $WW$ or $ZZ$ form two. The latter provides the highest sensitivity to the benchmark signals described in Section~\ref{sec:signalsamples} while the others probe the individual decay channels. Only the boson-tagging requirement differs between the regions. The highest mass jet of the two highest \pt{} jets is considered as the candidate for the higher boson mass requirement. For the $WZ$ selection this means that the highest mass jet must satisfy the $Z$ boson selections and the second highest one the $W$ boson selections. The selection requirements are summarised in Table~\ref{tab:selection}.
\begin{table}[h!]
\caption{Event selection requirements and definition of the different regions used in the
analysis. Different requirements are indicated for the highest-\pt{} (leading) jet with
index 1 and the second highest-\pt{} (subleading) jet with index 2.
}
\begin{center}
\begin{tabular}{l  l}
\hline\hline
Signal region      & Veto events with leptons: \\
& ~~~~~~ No $e$ or $\mu$ with $\pt > 25~\GeV{}$ and $|\eta| < 2.5$ \\
& Event pre-selection: \\
& ~~~~~~ $\geq 2$ large-$R$ jets with $|\eta| < 2.0$ and mass $> 50~\GeV{}$ \\
& ~~~~~~ $\ptone > 500~\GeV{}$ and $\pttwo > 200~\GeV{}$ \\
& ~~~~~~ $\mjj > \minmjj~\TeV{}$ \\
& Topology and boson tag: \\
& ~~~~~~ $|\Delta y| = |y_1 - y_2| < 1.2$ \\
& ~~~~~~ $A = (\ptone - \pttwo) \, / \, (\ptone + \pttwo) < 0.15$ \\
& ~~~~~~ Boson tag with \d2{} variable, \ntrk{} variable, and $W$ or $Z$ mass window \\
\midrule\midrule
$V$+jets control region
& Veto events with leptons: \\
& ~~~~~~ No $e$ or $\mu$ with $\pt > 25~\GeV{}$ and $|\eta| < 2.5$ \\
& $V$+jets selection: \\
& ~~~~~~ $\geq 2$ large-$R$ jets with $|\eta| < 2.0$ \\
& ~~~~~~ $\ptone > 600~\GeV{}$ and $\pttwo> 200~\GeV{}$ \\
& ~~~~~~ Boson tag with \d2{} and \ntrk{} variables on either jet \\
& ~~~~~~ Anti-boson tag with \d2{} variable on other jet \\
\hline\hline
\end{tabular}
\label{tab:selection}
\end{center}
\end{table}
 
The selection efficiency, defined as the number of selected events at different stages of the selection divided by the number of generated events, as a function of the resonance mass, is shown in Figure~\ref{objects:fig:efficiency} for the HVT $Z'$ decaying into $WW$ and for the bulk \grav{} decaying into $ZZ$.
Similar efficiencies are obtained in the $WZ$ final state for the
HVT model and in the $WW$ final state for the bulk RS models.
The figure shows that, among the different selection criteria described above,
the boson tagging reduces the signal efficiency the most. However, this particular
selection also provides the most significant suppression of the dominant
multijet background. The resulting width of the \mjj{} distributions in the signal region for a HVT model A $W^\prime\to WZ$ (Bulk RS graviton $\to ZZ$) is about 6\% (10\%) of its mean value across the mass range studied, corresponding to about $120~\GeV{}$ ($200~\GeV{}$) at $2~\TeV{}$.
Multijet background events are suppressed with a rejection factor greater than $10^5$ across the entire \mjj{} search range, as determined from simulation.
 
\begin{figure}[htp!]
\begin{center}
\subfloat[]{
\includegraphics[width=0.45\textwidth]{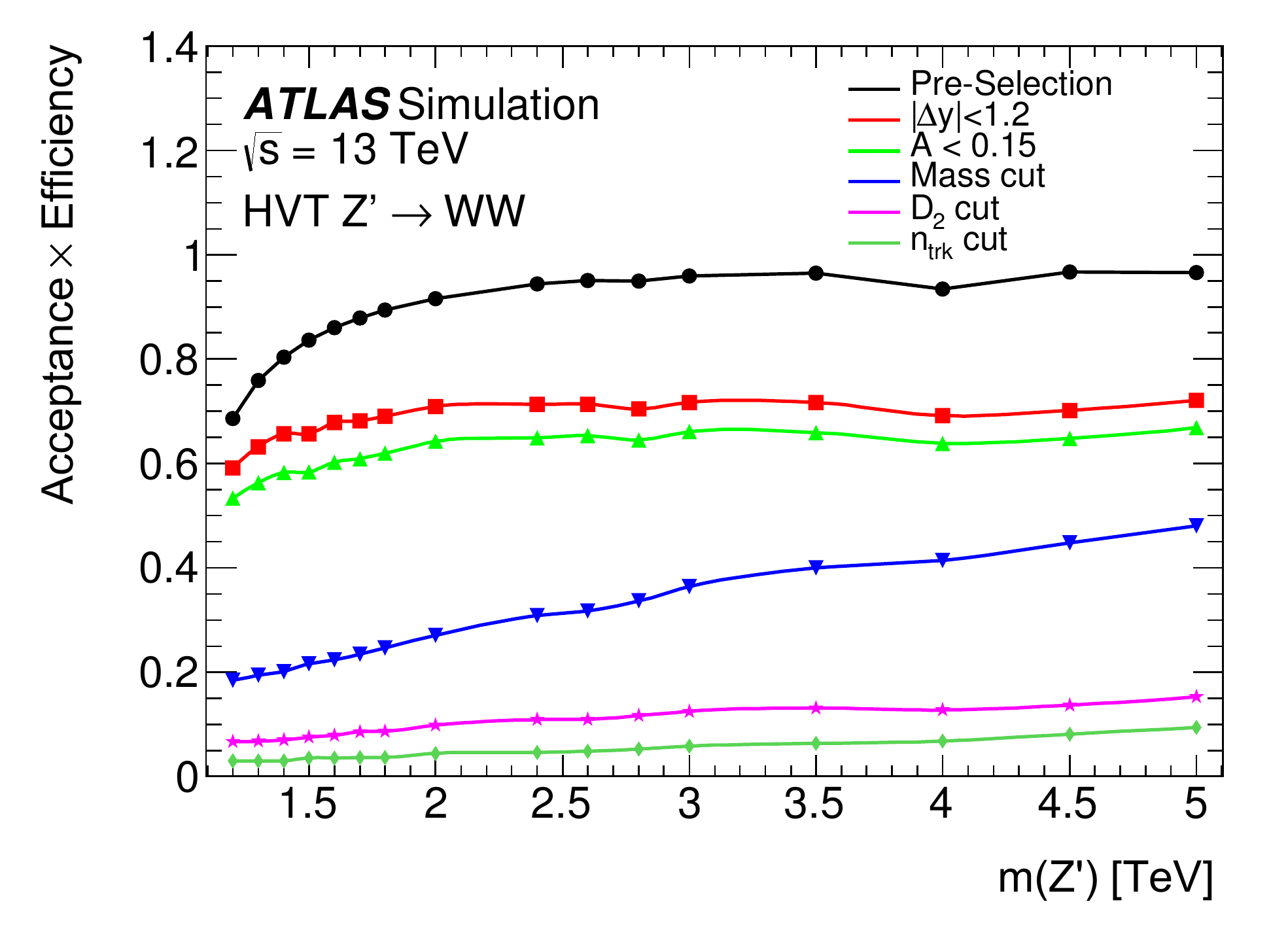}\label{fig:eff_hvt_ww}
}
\subfloat[]{
\includegraphics[width=0.45\textwidth]{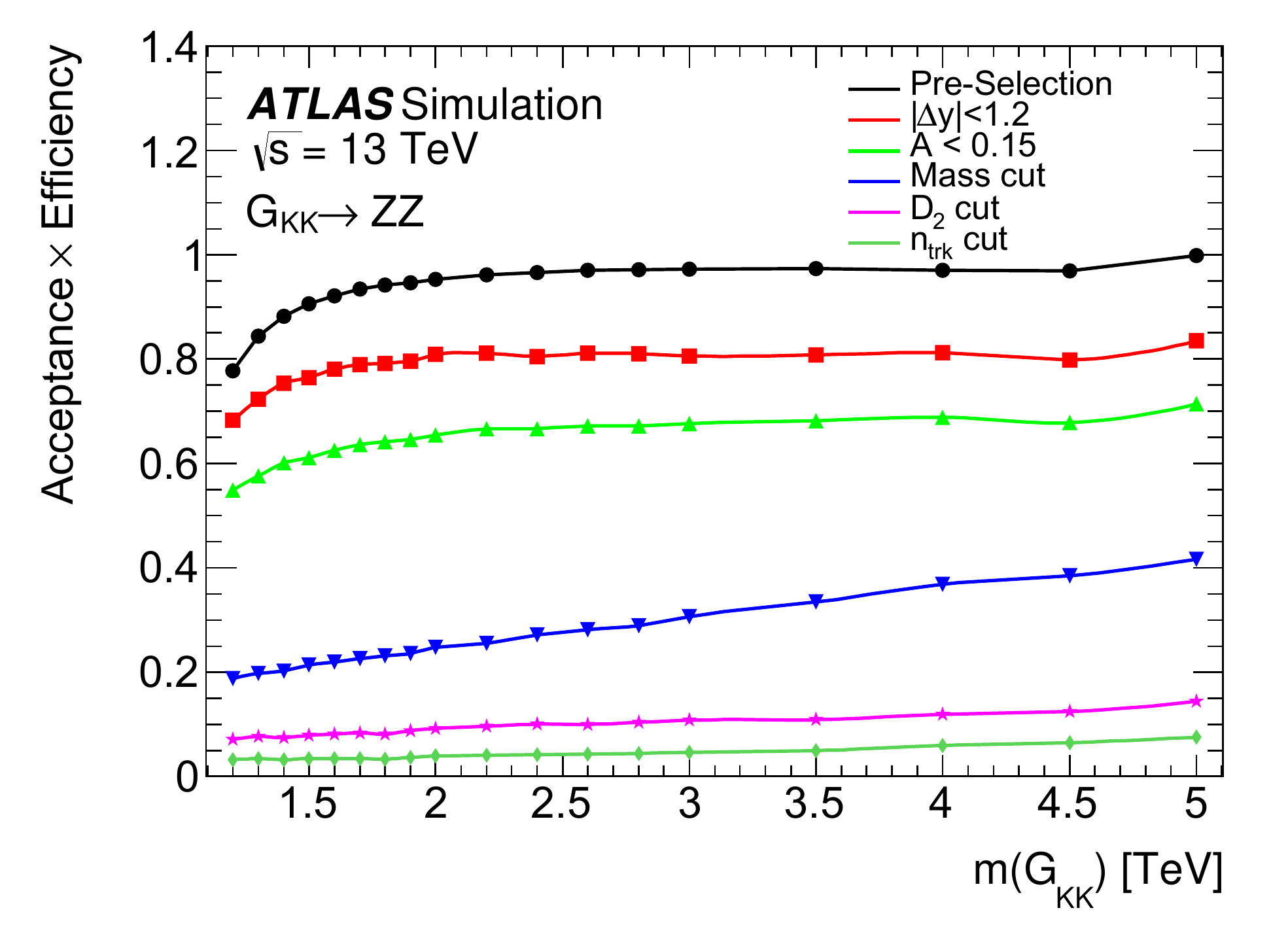}\label{fig:eff_gzz}
}
\end{center}
\vspace{-20 pt}
\caption{ The acceptance $\times$ efficiency for the selection, defined as the number of selected events at
different stages of the selection divided by the number of generated events, for
(a) HVT $Z^{\prime}\ra WW$ and (b) $\grav \ra ZZ$ as a
function of mass.
The selections are applied in sequence and include pre-selections,
topological selections on \dy{}, \pt{} asymmetry $A$, and boson tagging using
jet mass, \d2{}, and \ntrk{}. }
\label{objects:fig:efficiency}
\end{figure}
\section{Background parameterisation}
\label{sec:background}
 
The search for diboson resonances is performed by looking for narrow peaks
above the smoothly falling \mjj{} distribution expected from the SM.
The background in the search is estimated empirically from the observed \mjj{} spectrum in the signal region. The background estimation procedure is based on a binned maximum-likelihood
fit of the following parameterised form to the observed \mjj{} spectrum:
 
\begin{equation}
\label{eqn:dijet12}
\frac{\mathrm{d}n}{\mathrm{d}x} = p_1 (1-x)^{p_2-\xi p_3} x^{-p_3}
\end{equation}
where $x = \mjj{}/\sqrt{s}$, $p_1$ is a normalisation factor, $p_2$ and $p_3$ are dimensionless shape parameters, and
$\xi$ is a constant. The value of $\xi$ is derived in an iterative way, minimising the correlation between $p_2$ and $p_3$ in the fit, for each \mjj{} distribution. It is confirmed that the complexity of this fit function is sufficient for the expected number of events in the signal regions by performing Wilks likelihood-ratio tests~\cite{Wilks:1938dza}.
The fit is performed to the \mjj{} distribution in each signal region in data
with a constant bin size of $100~\GeV{}$. This choice is motivated by the experimental resolution.
 
The modelling of the parametric shape in Eq~\eqref{eqn:dijet12} is tested in dedicated fit control regions (CRs) in data. These CRs are designed to resemble the expected background in the SRs in both their shape and number of events, assuming that no signal contribution is present. Four regions are defined as shown in Figure~\ref{fig:ABCDSketch}, where A and B differ for each tested SR. A possible contamination in region A, C, or D from a potential beyond-the-SM signal is negligible. Region B corresponds to the nominal signal regions.
 
\begin{figure}[hpt!]
\begin{center}
\includegraphics[width=0.3\textwidth]{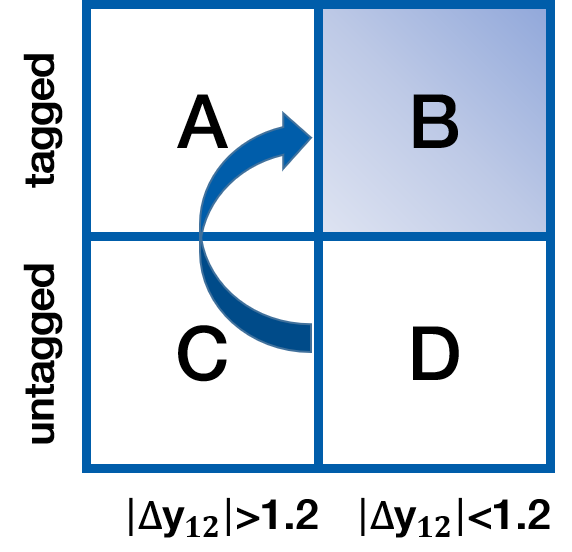}
\end{center}
\caption{Four orthogonal regions used to build the fit control region for each signal region. A: $\dy{}>1.2$ with both the jets boson-tagged, B: $\dy{}<1.2$ with both the jets boson-tagged (this is the nominal signal region), C: $\dy{}>1.2$ with the event not boson-tagged, D: $\dy{}<1.2$ with the event not boson-tagged. Regions A and C are used to derive a per-event transfer factor from region D to the fit control region, which is representative of region B. A and C are also signal-depleted due to the $\dy{}>1.2$ requirement.
}
\label{fig:ABCDSketch}
\end{figure}
 
The probability of misidentifying either the highest or second highest mass jet in an event as a $W$ or $Z$ boson in a data sample dominated by multijets is parameterised  as a function of jet \pt{} using regions C and A. It is validated on data that such a probability is independent of \dy{}. Since a misidentification correlation between the two leading jets of the multijet background is observed after the pre-selections, the probability of the second highest mass jet is derived by requiring the highest mass jet to be in the mass window of the boson tagger. By applying per-jet weights, for the inverted selections, depending on the jet \pt{}, events of the region D are transformed to resemble region B -- the fit CRs. To correctly take into account the expected statistical fluctuations and uncertainties, the CR distributions are assigned the correct Poisson errors, and fluctuated accordingly. The last step is repeated multiple times, fitting each distribution with the background fit function, and evaluating the goodness-of-fit $\chi^2/$NDF. Bins with fewer than five events are combined with bins that contain at least five events to compute the number of degrees of freedom (NDF). On average, the $\chi^2/$NDF is equal to unity with no cases for which the fit fails.
Figure~\ref{fig:CRfullDS} shows the fit result performed in an example $WZ$ fit CR.  Similar results are obtained for the other CRs, confirming the ability of the chosen background fit function (Eq.~\eqref{eqn:dijet12}) to describe the expected background dijet mass spectra in the SRs. It is validated on both the data and the simulation that this parametric background description is valid up to $8.0~\TeV{}$, which is also the mass up to which the observed \mjj{} spectra are fit.
 
\begin{figure}[hpt!]
\begin{center}
\includegraphics[width=0.5\textwidth]{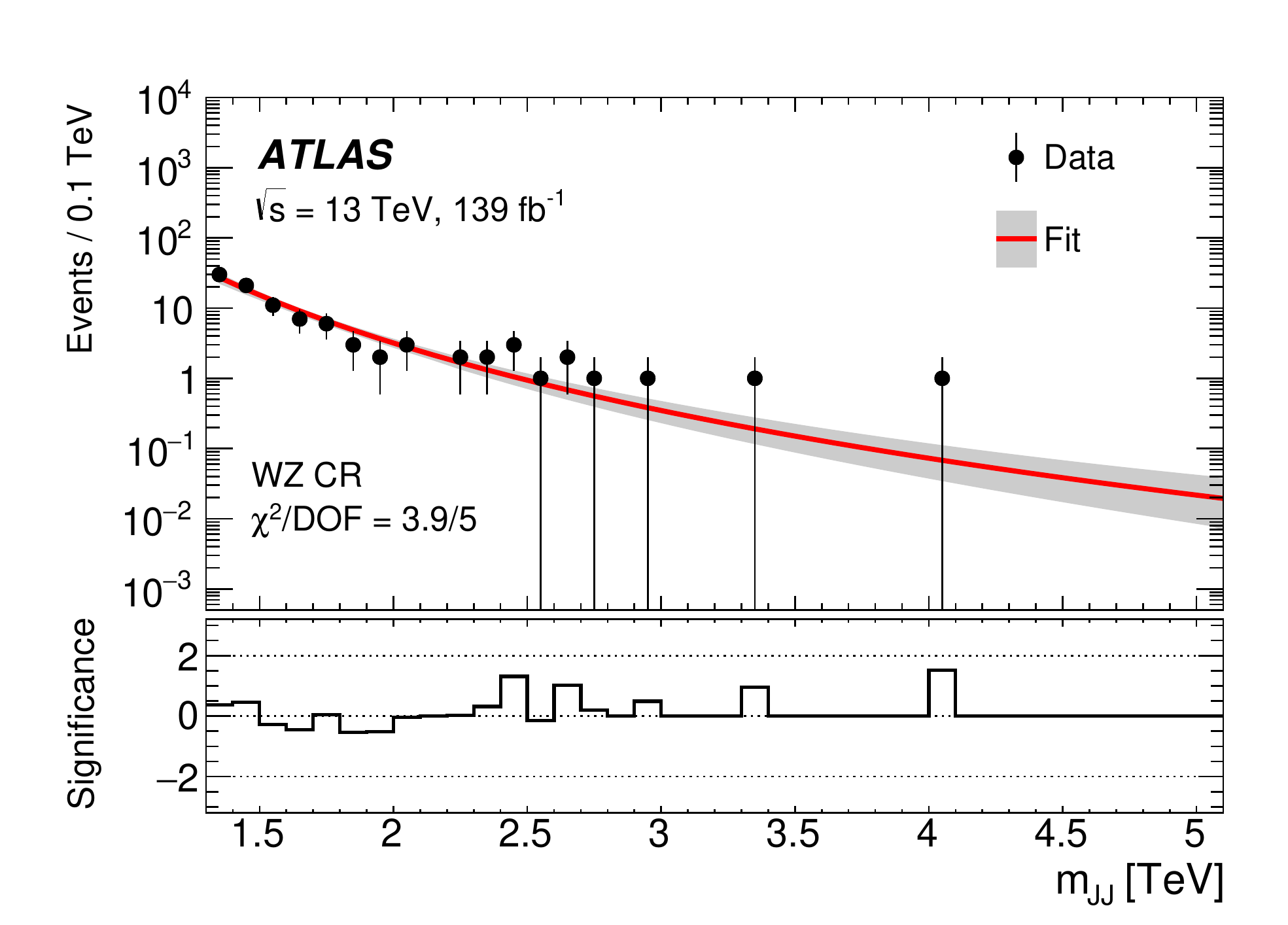}\label{fig:WZCR}
\end{center}
\caption{Comparison between fitted background shape and the \mjj{} spectra in an example $WZ$ fit control region in data.
The fitted background distribution is normalised to the data shown in the displayed mass range.
The shaded bands represent the uncertainty in the background expectation calculated from the maximum-likelihood function.
The lower panel shows the significance, defined as the $z$-value as described in Ref.~\cite{Choudalkis:2012}.
}
\label{fig:CRfullDS}
\end{figure}
The statistical uncertainty in the background expectation comes directly from the uncertainty in the fitted parameters of the background function,
which assumes a smoothly falling \mjj{} distribution.
Possible additional uncertainties due to the background model are assessed by considering signal-plus-background fits (also called spurious signal tests) of the chosen function to the fit control regions of data in which a signal contribution is expected to be negligible.
The background is modelled with Eq.~\eqref{eqn:dijet12} and the signal is modelled using resonance mass distributions from simulation.
These procedures were estimated to introduce a bias smaller than 25\% of the statistical uncertainty in the background estimate at any mass in the search region, and no additional uncertainty is assigned.
\section{Systematic uncertainties}
\label{sec:systematics}
The uncertainties affecting the background modelling are taken directly from the errors in the fit parameters of the background
estimation procedure described in Section~\ref{sec:background}.
The systematic uncertainties in the expected signal yield and shapes arise from detector effects and MC modelling and are assessed and expressed in terms of nuisance parameters
in the statistical analysis as described in Section~\ref{sec:statistics}.
The dominant sources of uncertainty in the signal modelling arise from uncertainties in the large-$R$ jet tagging efficiency and the jet \pt{} calibration. These two uncertainties, and the uncertainty in the fitted background, are also the only ones significantly affecting the statistical results.
 
The uncertainty in the jet \pt{} scale (J$\pt$S) is evaluated using track-to-calorimeter double ratios between data and simulation~\cite{PERF-2012-02}.
The ratio of the calorimeter and track measures of jet \pt{} is expected to be the same in data and simulation and any observed differences are assigned as baseline systematic uncertainties. Uncertainties obtained from this procedure assume no correlation between the two \pt{} measures, while any residual correlation would modify them by a certain factor.
An upper limit to the correlation between the two \pt{} measures is found to be at the percent level by comparing the results of this double-ratio procedure between jets built from TCC inputs and jets built from calorimeter-only inputs.
Additional uncertainties due to the track reconstruction efficiency,
track impact parameter resolution, and track fake rate are taken into account.
The size of the total J$\pt$S uncertainty varies with jet \pt{} and
is between $2.5\%$ and $5\%$ for the full mass range.
 
The impact of the jet \pt{} resolution uncertainty is evaluated
event-by-event by rerunning the analysis with an additional Gaussian smearing applied to the input jets' \pt{} to degrade the nominal resolution by the systematic uncertainty value.
The systematic uncertainty in the width of the Gaussian distribution is an absolute 2\% per jet, and is symmetrised.
 
Uncertainty in the jet mass scale and resolution influences the observed jet mass, affecting
the boson-tagging efficiency.
Any uncertainty in the value of the boson-tagging discriminant \d2{} or \ntrk{}, would also affect the selection efficiency of the analysis. A scale-factor for the $W/Z$-tagging efficiency is derived as described in Section~\ref{sec:vplusjets}.
The changes to the overall yield is hence corrected by $0.85_{-0.21}^{+0.23}$ per event with the boson-tagging efficiency scale-factor, assuming full correlation between the two jets. The uncertainty in the scale-factor is assigned as a two-sided variation in the yield. Additional studies comparing jet properties in data and simulation confirm this uncertainty to be valid up to $\maxmjj~\TeV{}$.
 
The uncertainty in the combined 2015--2018 integrated luminosity is 1.7\%~\cite{ATLAS-CONF-2019-021}, obtained using the LUCID-2 detector~\cite{LUCID2} for the primary luminosity measurements. The uncertainty from the trigger selection is found to be negligible, as the minimum requirement on the dijet invariant mass of $\minmjj~\TeV{}$ guarantees that the trigger is fully efficient.

Uncertainties in the behaviour of the PDFs at high $Q^2$ values
can potentially have a large effect on the signal acceptance.
This systematic uncertainty is estimated by taking the envelope formed by the largest deviations produced by the errors of three PDF sets, as set out by the PDF4LHC group~\cite{Butterworth:2015oua}.
A constant 1\% uncertainty is applied in the case of the RS and radion models, and a pole-mass-dependent uncertainty ranging from 1\%--12\% is applied in the case of the HVT model.
Systematic variations are used to cover uncertainties in the A14 tuned parameter values describing
initial-state radiation, final-state radiation, and multi-parton interactions.
The uncertainty in the signal acceptance is evaluated at the generator level, before boson-tagging requirements. Following the same procedure as for the PDFs, constant uncertainties of 3\% (5\%) are applied for the HVT (RS and radion) models.
\section{Results}
\label{sec:results}
\subsection{Background fit}
\label{sec:Fit}
Figure~\ref{Res:fig:mJJ2} shows the comparison of the dijet mass distributions of the selected events in the combined $WW+WZ$ and $WW+ZZ$ signal regions with the expected background
distribution from the background-only fits to the data. The fitted background functions shown, labelled `Fit', are evaluated in bins between $\minmjj~\TeV{}$ and $8.0~\TeV{}$. No events are observed beyond $5.0~\TeV{}$.
A total of 119 and 113 events are observed above $\minmjj~\TeV{}$ in the $WW+WZ$ and $WW+ZZ$ signal regions, respectively.
Due to the non-exclusive selections of the boson taggers, about 50\% of events satisfying the $WW$ selection also satisfy the $ZZ$ selection. The highest mass event at $4.4~\TeV{}$ is the same for both signal regions, and it is compatible with the background expectation in the high mass region.
 
\begin{figure}[th!]
\begin{center}
\subfloat[]{
\includegraphics[width=0.45\textwidth]{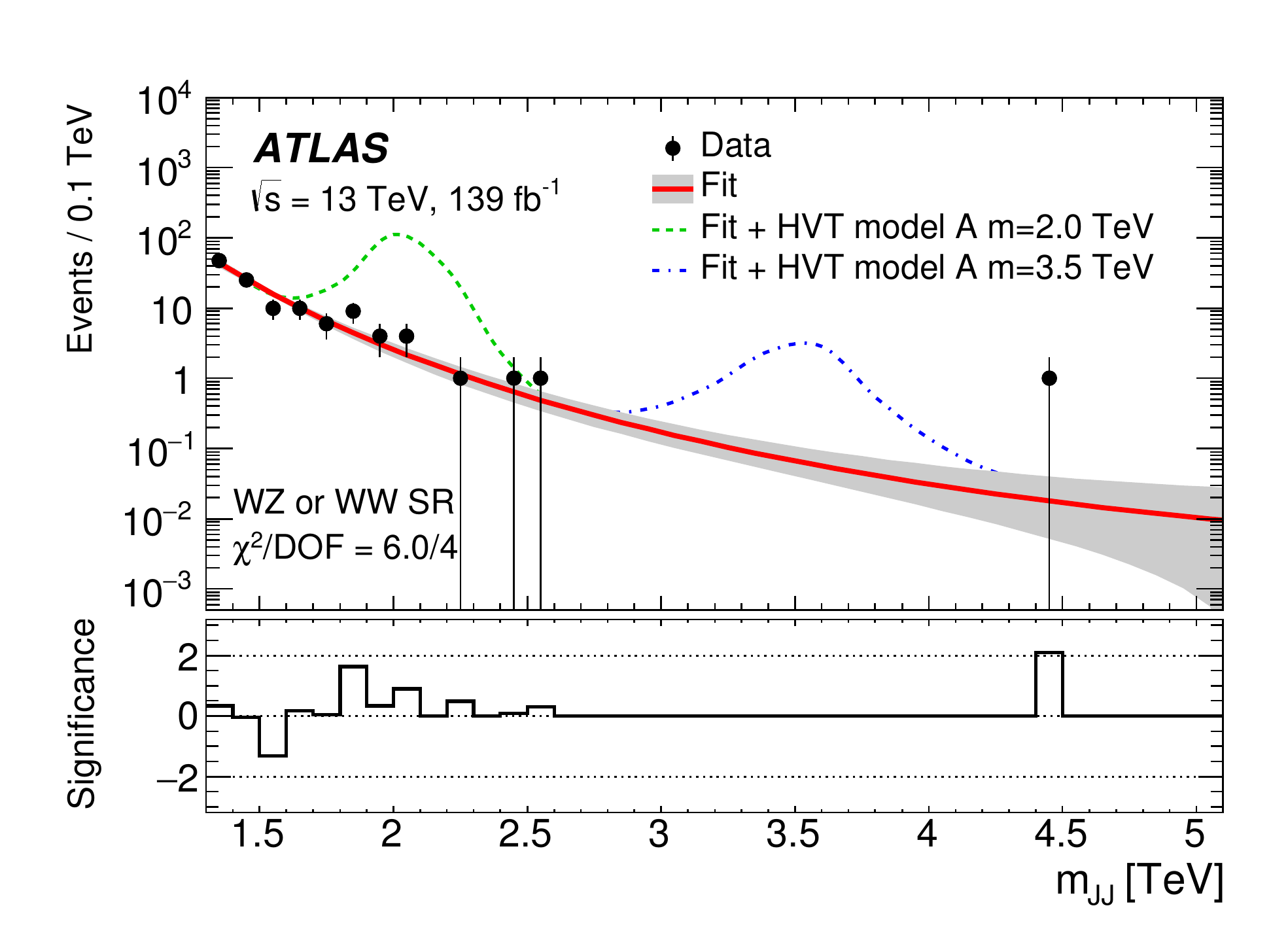}\label{Res:fig:WZorWW}
}
\subfloat[]{
\includegraphics[width=0.45\textwidth]{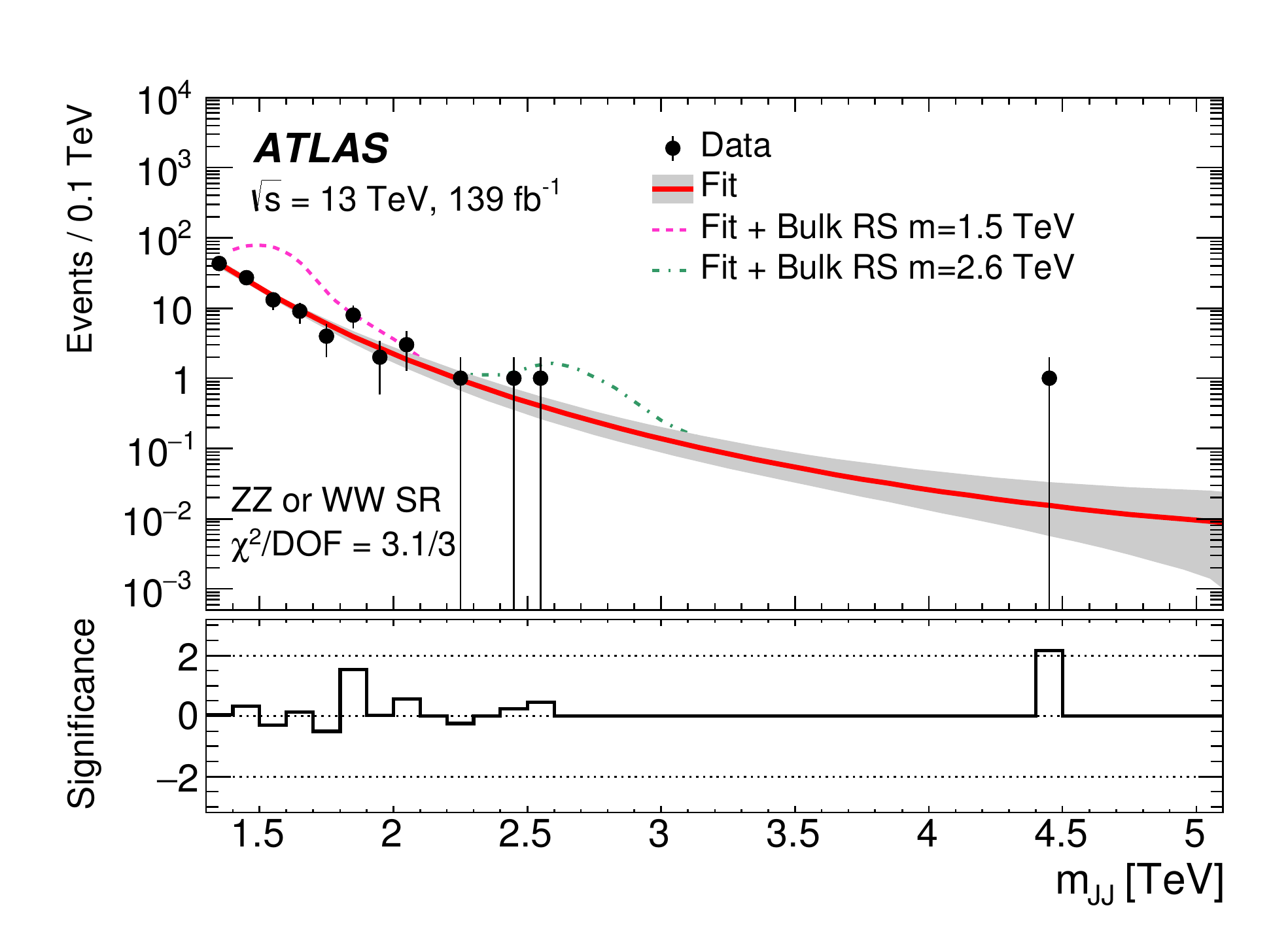}\label{Res:fig:WWorZZ}
}
\end{center}
\caption{Background-only fits to the dijet mass (\mjj{}) distributions in data after tagging in the combined (a)~$WW+WZ$, and (b)~$WW+ZZ$ signal region. The shaded
bands represent the uncertainty in the background expectation calculated from the maximum-likelihood function. The lower panels show the significance, defined as the $z$-value as described in Ref. [68]. Selected theoretical signal distributions are overlaid on top of the background.}
\label{Res:fig:mJJ2}
\end{figure}
 
\subsection{Statistical analysis}
\label{sec:statistics}
In the statistical analysis, the parameter of interest is the \textit{signal strength}, which is defined as
a scale-factor to the predicted signal normalisation of the model being tested. The analysis follows the \textit{frequentist} approach with a test statistic based on the
profile-likelihood ratio~\cite{Cowan:2010js}. The test statistic extracts information about the signal strength from the binned
maximum-likelihood fit of the signal-plus-background model to the data.
The likelihood model is defined as,
\begin{equation*}
\label{eqn:lhood}
{\cal L} = \prod_i P_{\mathrm{pois}}( \nobsi | \nexpi ) \times G(\alpha) \times {\cal N}(\mathbf{\theta})
\end{equation*}
where $P_{\mathrm{pois}}( \nobsi | \nexpi )$ is the Poisson probability to observe $\nobsi$ events
if $\nexpi$ events are expected, $G(\alpha)$ are a series of Gaussian probability density functions modelling the systematic uncertainties, $\alpha$, related to the shape of the signal, and ${\cal N}(\mathbf{\theta})$ is a log-normal distribution for the nuisance parameters, $\mathbf{\theta}$,  modelling the systematic uncertainty in the signal normalisation.
The expected number of events is the bin-wise sum of those expected for the
signal and background:
$\nexpvec = \nsigvec + \nbgvec$.
The expected number of  background events in dijet mass bin $i$, \nbgi{},
is obtained by integrating $\mathrm{d}n/\mathrm{d}x$
obtained from Eq.~\eqref{eqn:dijet12} over that bin.
Thus \nbgvec{} is a function of the dijet background
parameters $p_1$, $p_2$ and $p_3$. The expected number of signal  events, \nsigvec{},
is evaluated from MC simulation assuming the cross-section of the model under test multiplied by the signal strength,
including the effects of the systematic uncertainties described in Section~\ref{sec:systematics}.
 
The significance of observed excesses over the background-only prediction is quantified using the local $p_{0}$-value, defined as the probability of the background-only model to produce a signal-like fluctuation at
least as large as that observed in the data. The most extreme $p_0$ has a local significance of 1.8 standard
deviations, and is found when testing the HVT $W' \ra WW$ hypothesis at a resonance mass of $1.8~\TeV{}$.
This is within the expected fluctuation of the background.
 
Limits at 95\% confidence level (CL) on the production cross-section times branching fraction to diboson final states for the benchmark signals are set with sampling distributions generated using pseudo-experiments.
All systematic uncertainties are considered. The uncertainty in the $W/Z$-tagging efficiency is dominant at lower masses, while the uncertainty in the background modelling has largest impact at high masses. Uncertainties in the jet \pt{} scale are at the percent level but are subordinate across the full mass range. The cross-section limits extracted for the different benchmark scenarios in the $WW+WZ$ and  $WW+ZZ$ signal regions are shown in Figure~\ref{fig:limitsHVT1} and Table~\ref{Res:tab:radionlim}. Table~\ref{Res:tab:excl_mass} presents the resonance mass ranges excluded at the 95\% CL in the various signal regions and signal models considered in the search.
 
\begin{figure}[th!]
\begin{center}
\subfloat[]
{
\includegraphics[width=0.45\textwidth]{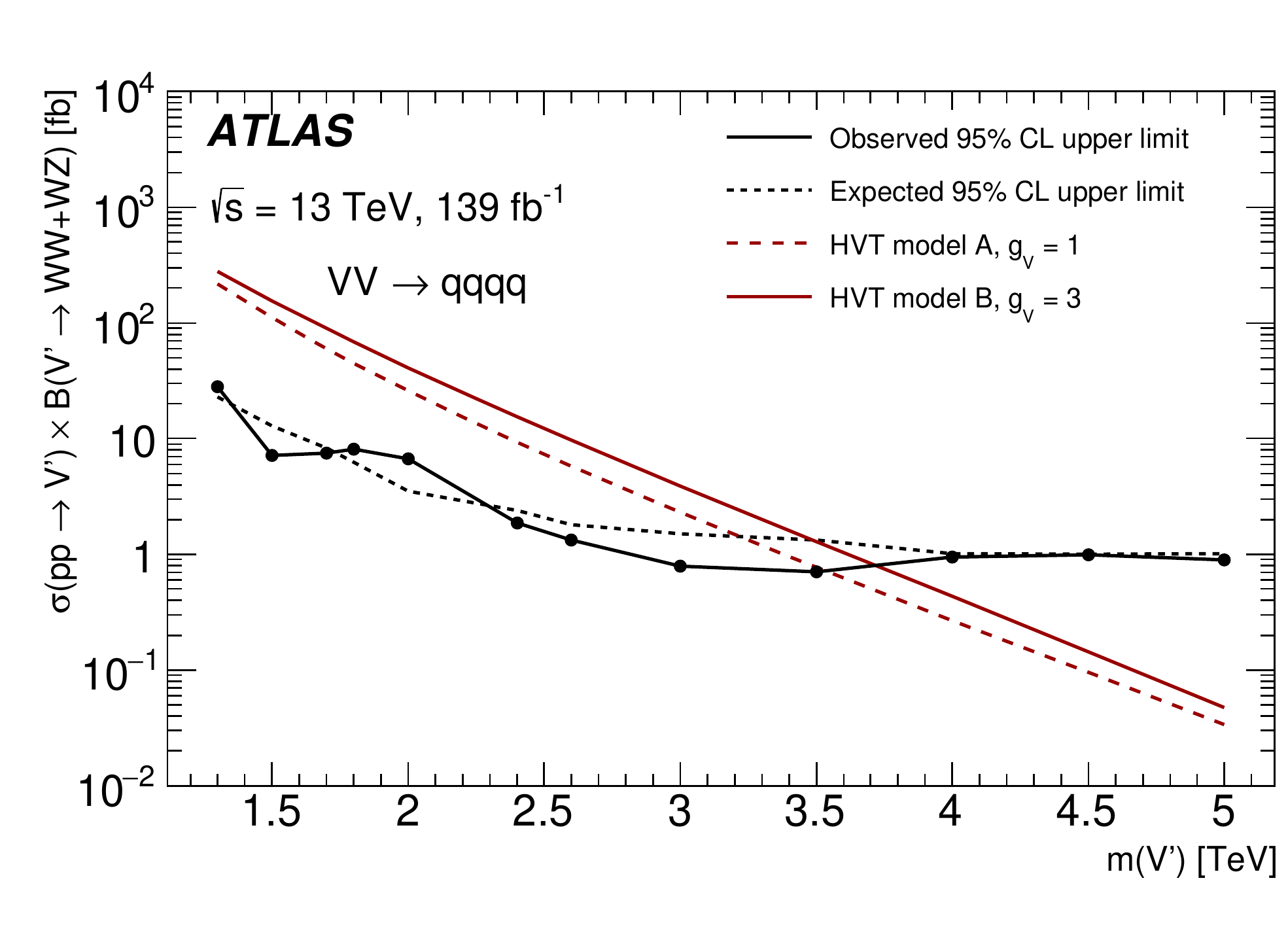}
\label{fig:limitsHVT}
}
\subfloat[]
{
\includegraphics[width=0.45\textwidth]{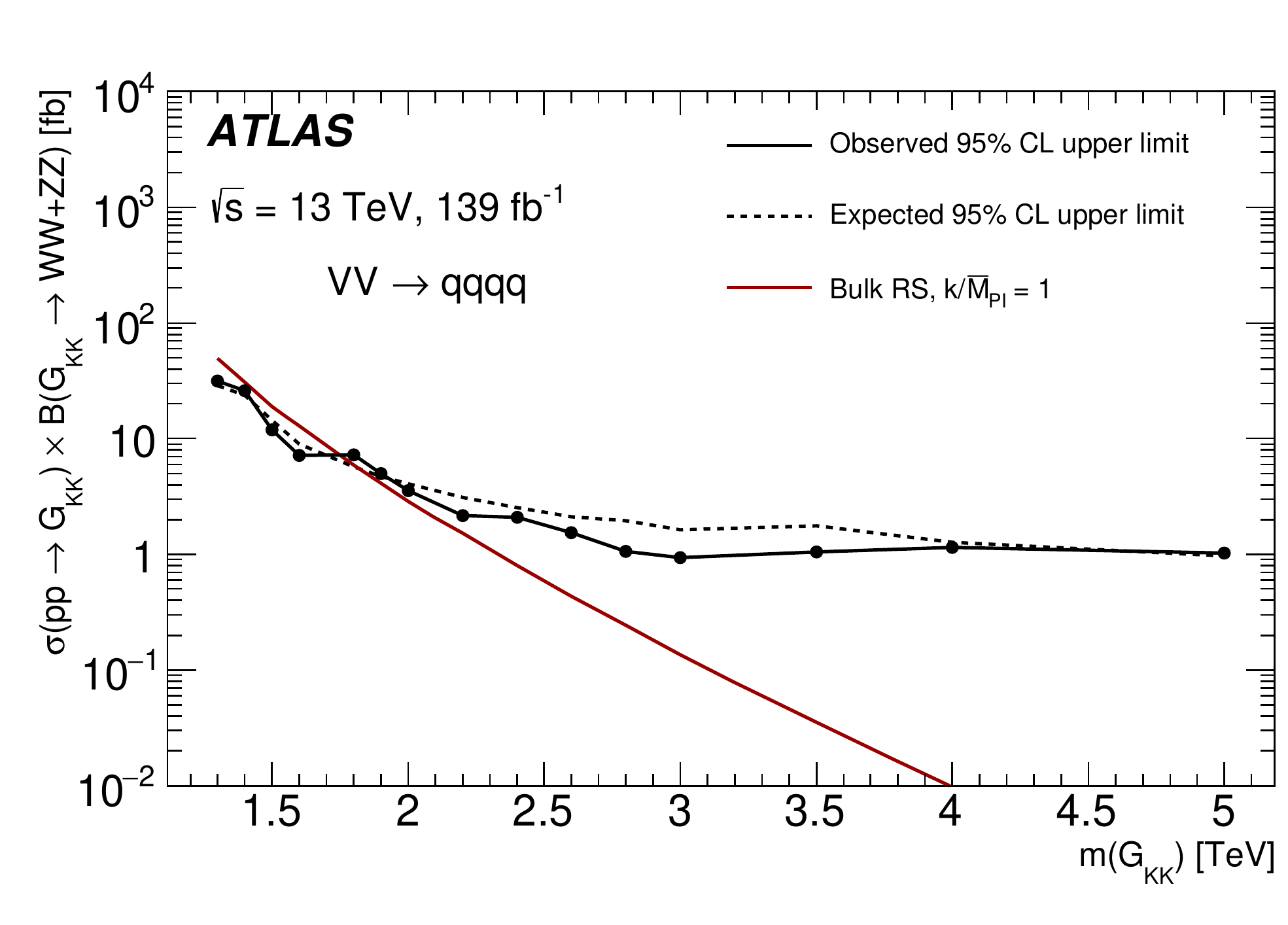}
\label{fig:limitsRSG}
}
\end{center}
\vspace{-20 pt}
\caption{Observed and expected limits at 95\% CL on the cross-section times
branching ratio for $WW+WZ$ production as a function of (a) $m_\mathrm{V^\prime}$, and for $WW+ZZ$ production as a function of (b) the Bulk RS graviton $m_\mathrm{\grav}$. The predicted cross-section times branching ratio is shown (a) as dashed and solid lines for the HVT models A with $g_V = 1$ and B with $g_V = 3$, respectively, and (b) as a solid line for the bulk RS model with $k/\overline{M}_{\mathrm{Pl}} = 1$.}
\label{fig:limitsHVT1}
\end{figure}
 
\begin{table}[tpb!]
\caption{Observed and expected limits at 95\% CL on cross-section times
branching ratio for $WW+ZZ$ production for different radion masses $m_\mathrm{\radion}$, as well as the predicted cross-section times branching ratio.
\label{Res:tab:radionlim}
}
\centering
\begin{tabular}{SScS}
\hline
\hline
\multicolumn{1}{c}{Mass [\TeV\unskip]} & \multicolumn{1}{c}{Observed Limit [fb]} & Expected Limit [fb] & \multicolumn{1}{c}{Prediction [fb]} \\
\hline
2.0      & 5.72   &     $5.75$ & 9.4 \\
3.0      & 1.86   &     $2.85$ & 0.92 \\
4.0      & 1.98   &     $2.34$ & 0.089 \\
5.0      & 1.98   &     $2.02$ & 0.012 \\
\hline
\hline
\end{tabular}
\end{table}
 
\begin{table}[tpb!]
\caption{Observed excluded resonance masses (at 95\% CL) in the individual and combined signal regions for the HVT, bulk RS and radion models.
\label{Res:tab:excl_mass}
}
\begin{center}
\begin{tabular}{ccc}
\hline
\hline
Model   & Signal Region & Excluded mass range [$\TeV{}$]\\
\hline
& $WW$ & $1.3$--$2.8$ \\
Radion             & $ZZ$ &  $1.6$--$2.1$   \\
& $WW+ZZ$ & $\radionlimLow$--$\radionlimHigh$ \\
\hline
& $WW$ & $1.3$--$2.9$ \\
HVT model A, $g_V = 1$ & $WZ$ & $1.3$--$3.4$ \\
& $WW+WZ$ & $\hvtAlimLow$--$\hvtAlimHigh$ \\
\hline
& $WW$ & $1.3$--$3.1$ \\
HVT model B, $g_V = 3$ & $WZ$ & $1.3$--$3.6$ \\
& $WW+WZ$ & $\hvtBlimLow$--$\hvtBlimHigh$ \\
\hline
& $WW$ & $1.3$--$1.6$ \\
Bulk RS, $k/\overline{M}_{\mathrm{Pl}} = 1$ & $ZZ$ &  none  \\
& $WW+ZZ$ & $\gravlimLow$--$\gravlimHigh$ \\
\hline
\hline
\end{tabular}
\end{center}
\end{table}
\FloatBarrier
\section{Conclusion}
\label{sec:conclusions}
A search for narrow heavy resonances decaying into dibosons in the
all hadronic channel is performed using $\lumi~\ifb$ of proton--proton collisions at $\sqrt{s} = 13~\TeV{}$
collected by the ATLAS experiment at the LHC from 2015 to 2018.
The results of the search are shown for the $WW+WZ$, $WW+ZZ$ channels, and are interpreted
in terms of a radion model, two HVT benchmark models, and a bulk \grav{} model.
The data are in  agreement with the background expectations in all channels.
Upper limits on the production cross-section times branching ratio to diboson final states for
new resonances with masses greater than $\minmjj~\TeV{}$ are set at the 95\% CL.
These results exclude at the 95\% CL the production of $WW+WZ$ from the HVT model A (model B) with $g_V = 1$ ($g_V = 3$)
with masses in the range of
$\hvtAlimLow~\TeV{}$--$\hvtAlimHigh~\TeV{}$ ~($\hvtBlimLow~\TeV{}$--$\hvtBlimHigh~\TeV{}$).
Production of a $\grav$ in the bulk RS model with $k/\overline{M}_{\mathrm{Pl}} = 1$ is excluded in the range $\gravlimLow~\TeV{}$--$\gravlimHigh~\TeV{}$, at the $95\%$ CL.
The production of $WW+ZZ$ from the scalar-like radion is excluded at the 95\% CL in the range $\radionlimLow~\TeV{}$--$\radionlimHigh~\TeV{}$.
 
\section*{Acknowledgements}
 

We thank CERN for the very successful operation of the LHC, as well as the
support staff from our institutions without whom ATLAS could not be
operated efficiently.
 
We acknowledge the support of ANPCyT, Argentina; YerPhI, Armenia; ARC, Australia; BMWFW and FWF, Austria; ANAS, Azerbaijan; SSTC, Belarus; CNPq and FAPESP, Brazil; NSERC, NRC and CFI, Canada; CERN; CONICYT, Chile; CAS, MOST and NSFC, China; COLCIENCIAS, Colombia; MSMT CR, MPO CR and VSC CR, Czech Republic; DNRF and DNSRC, Denmark; IN2P3-CNRS and CEA-DRF/IRFU, France; SRNSFG, Georgia; BMBF, HGF and MPG, Germany; GSRT, Greece; RGC and Hong Kong SAR, China; ISF and Benoziyo Center, Israel; INFN, Italy; MEXT and JSPS, Japan; CNRST, Morocco; NWO, Netherlands; RCN, Norway; MNiSW and NCN, Poland; FCT, Portugal; MNE/IFA, Romania; MES of Russia and NRC KI, Russia Federation; JINR; MESTD, Serbia; MSSR, Slovakia; ARRS and MIZ\v{S}, Slovenia; DST/NRF, South Africa; MINECO, Spain; SRC and Wallenberg Foundation, Sweden; SERI, SNSF and Cantons of Bern and Geneva, Switzerland; MOST, Taiwan; TAEK, Turkey; STFC, United Kingdom; DOE and NSF, United States of America. In addition, individual groups and members have received support from BCKDF, CANARIE, Compute Canada and CRC, Canada; ERC, ERDF, Horizon 2020, Marie Sk{\l}odowska-Curie Actions and COST, European Union; Investissements d'Avenir Labex, Investissements d'Avenir Idex and ANR, France; DFG and AvH Foundation, Germany; Herakleitos, Thales and Aristeia programmes co-financed by EU-ESF and the Greek NSRF, Greece; BSF-NSF and GIF, Israel; CERCA Programme Generalitat de Catalunya and PROMETEO Programme Generalitat Valenciana, Spain; G\"{o}ran Gustafssons Stiftelse, Sweden; The Royal Society and Leverhulme Trust, United Kingdom.
 
The crucial computing support from all WLCG partners is acknowledged gratefully, in particular from CERN, the ATLAS Tier-1 facilities at TRIUMF (Canada), NDGF (Denmark, Norway, Sweden), CC-IN2P3 (France), KIT/GridKA (Germany), INFN-CNAF (Italy), NL-T1 (Netherlands), PIC (Spain), ASGC (Taiwan), RAL (UK) and BNL (USA), the Tier-2 facilities worldwide and large non-WLCG resource providers. Major contributors of computing resources are listed in Ref.~\cite{ATL-GEN-PUB-2016-002}.
 

\printbibliography
\FloatBarrier
\clearpage
 
\begin{flushleft}
{\Large The ATLAS Collaboration}

\bigskip

G.~Aad$^\textrm{\scriptsize 101}$,    
B.~Abbott$^\textrm{\scriptsize 128}$,    
D.C.~Abbott$^\textrm{\scriptsize 102}$,    
O.~Abdinov$^\textrm{\scriptsize 13,*}$,    
A.~Abed~Abud$^\textrm{\scriptsize 70a,70b}$,    
K.~Abeling$^\textrm{\scriptsize 53}$,    
D.K.~Abhayasinghe$^\textrm{\scriptsize 93}$,    
S.H.~Abidi$^\textrm{\scriptsize 167}$,    
O.S.~AbouZeid$^\textrm{\scriptsize 40}$,    
N.L.~Abraham$^\textrm{\scriptsize 156}$
% ...
\bigskip
\\

% ...
$^{13}$Institute of Physics, Azerbaijan Academy of Sciences, Baku; Azerbaijan.\\
% ...
$^{40}$Niels Bohr Institute, University of Copenhagen, Copenhagen; Denmark.\\
% ...
$^{53}$II. Physikalisches Institut, Georg-August-Universit\"{a}t G\"ottingen, G\"ottingen; Germany.\\
% ...
$^{70}$$^{(a)}$INFN Sezione di Pavia;$^{(b)}$Dipartimento di Fisica, Universit\`a di Pavia, Pavia; Italy.\\
% ...
$^{93}$Department of Physics, Royal Holloway University of London, Egham; United Kingdom.\\
% ...
$^{101}$CPPM, Aix-Marseille Universit\'e, CNRS/IN2P3, Marseille; France.\\
$^{102}$Department of Physics, University of Massachusetts, Amherst MA; United States of America.\\
% ...
$^{128}$Homer L. Dodge Department of Physics and Astronomy, University of Oklahoma, Norman OK; United States of America.\\
% ...
$^{156}$Department of Physics and Astronomy, University of Sussex, Brighton; United Kingdom.\\
% ...
$^{167}$Department of Physics, University of Toronto, Toronto ON; Canada.\\
% ...
$^{a}$ Also at Centre for High Performance Computing, CSIR Campus, Rosebank, Cape Town; South Africa.\\
$^{b}$ Also at CERN, Geneva; Switzerland.\\
% ...
$^{*}$ Deceased

\end{flushleft}

\clearpage

\end{document}